\documentstyle[aps,prl,eqsecnum,multicol,epsfig]{revtex}

\renewcommand{\narrowtext}{\begin{multicols}{2} \global\columnwidth20.5pc}
\renewcommand{\widetext}{\end{multicols} \global\columnwidth42.5pc}
\newcommand{\Lrule}{\vspace*{-0.2in}\noindent\vrule width3.5in height.2pt
  depth.2pt \vrule depth0em height1em}
\newcommand{\Rrule}{\vspace{-0.1in}\hfill\vrule depth1em height0pt \vrule
  width3.5in height.2pt depth.2pt\vspace*{-0.1in}}

\begin{document}
\draft
\title{Limits of the dynamical approach to non-linear response of
mesoscopic systems}
\author{V.I.Yudson$^{1,2}$}
\address{$^{1}$
Center for Frontier Science, Chiba University,
1-33 Yayoi-cho, Inage-ku, Chiba 263-8522, Japan\\
$^{2}$ Institute of Spectroscopy, Russian Academy of Sciences,
Troitsk, Moscow region, 142090 Russia}
\author{E.Kanzieper$^{3,5}$ and V.E.Kravtsov$^{3,4}$}
\address{$^3$ The Abdus Salam International Centre for Theoretical
Physics,
P.O. Box 586, 34100 Trieste, Italy\\
$^4$ Landau Institute for Theoretical Physics, 2 Kosygina Street, 117940
Moscow, Russia\\ $^5$ Hitachi Cambridge Laboratory, Madingley Road,
Cambridge CB3 0HE, UK}
\maketitle
\begin{abstract}
We have considered the nonlinear response of mesoscopic systems of
non-interacting electrons to the time-dependent external field. 
In this consideration the inelastic processes have been neglected
and the electron thermalization occurs due to the electron exchange with
the reservoirs. We have demonstrated that the diagrammatic technique
based on the method of analytical continuation or on the Keldysh formalism
is capable to describe the heating automatically. The corresponding
diagrams contain a novel element, {\it the loose diffuson}. We have shown
the equivalence of such a diagrammatic technique to the solution to the
kinetic equation for the electron energy distribution function. We have
identified two classes of problems with different behavior under ac
pumping. In one class of problems (persistent current fluctuations,
Kubo conductance) the observable depends on the electron energy
distribution renormalized by heating. In another class of problems
(Landauer 
conductance) the observable is insensitive to heating and depends on the
temperature of
electron reservoirs. As examples of such problems we have considered in
detail 
the persistent current fluctuations under ac pumping and two types of
conductance measurements (Landauer
conductance and Kubo conductance) that behave differently under ac
pumping.

\end{abstract}

\narrowtext
\section{Introduction}

Recently there has been a considerable interest in non-equilibrium
mesoscopics. The effect of adiabatic charge pumping \cite{Thou} has been   
experimentally observed \cite{Marc} and discussed theoretically
\cite{SAA}.
Weak localization in a quantum dot under ac pumping  has been
theoretically studied \cite{VA}.
The non-equilibrium noise has been suggested \cite{ann,krav3} as a cause
of
both the low
temperature dephasing saturation \cite{moh1} and the anomalously large
ensemble averaged persistent current \cite{Levi}.
The results of Ref.\cite{krav3}
are based on the earlier works \cite{DC} on the ensemble averaged dc
current caused by the quantum Aharonov-Bohm rectification of the external
ac electric field. Without the Aharonov-Bohm magnetic flux the rectified
dc current or voltage (`photovoltaic effect') has zero ensemble average
but can exist in individual mesoscopic samples because of the specific
arrangement of impurities or irregularities in the dot's shape. This
effect
was suggested long ago \cite{FKh}
and reconsidered very recently for the case of the quantum dot
\cite{VAA}.

Theoretical description of all the effects listed above requires to go
beyond the linear response theory and
to consider the essentially non-linear response to the ac pump  field.
This raises a question on the `minimal model' for the adequate description
of nonlinear responses in mesoscopic systems. For the linear conductance 
the minimal model is the system of non-interacting electrons
with the impurity scattering and interaction with the external
electric field. Such a model does not explicitly contain dissipation. Yet
it allows to obtain a correct value of conductivity that is the key
quantity for the dissipation function. We will refer to a description
based on a  model of
that sort as the `dynamical approach'. In this approach the electron-electron
(e-e) and electron-phonon (e-ph) interaction is neglected and the stationary regime
under
external pumping is reached in an open system via the escape of `hot'
electrons into the massive leads playing a role of an electron bath.  

The question we address in this paper is this: To what extent this model
applies
to the nonlinear phenomena in mesoscopic systems and how one can see its
limitations through intrinsic inconsistencies and physical paradoxes.

Another important issue we address in this paper is how to describe
heating effects by the impurity diagammatic technique without explicitly solving the
kinetic equation. It turns out that heating can be described automatically by the
new class of
diagrams containing the `loose diffusons' with one free end. They are contrasted to
the
ordinary diffusons and cooperons which are connected in loops by the `Hikami boxes'   
and describe the effect of electron phase-coherence. Thus we show the way to
{\it separate} the heating and the dephasing effects on the level of the impurity
diagrammatic technique.

The paper is organized as follows. In Sec.II and Sec.III we discuss the
general
structure of the perturbation theory in the external ac field using the
method of analytical continuation and the Keldysh technique and describe
simple  
rules of the impurity diagrammatic technique in the time domain. In
Sec.III
we derive the diffusion propagators (`diffusons' and `cooperons') in the
external ac field at different boundary conditions.
Sec.IV is central for the paper. There we introduce the `loose diffusons' 
and demonstrate that evaluating the diagrams with the loose diffusons is
equivalent to the solution of the kinetic equation
for the electron energy distribution. We also discuss
the paradoxes connected with the loose diffusons in closed electronic
systems. As an example of the role of the loose diffusons we consider in
Sec.V the variance of the persistent
current fluctuations under ac pumping. For mesoscopic rings connected to
an electron reservoir by a passive lead we compute the temperature
dependence of the persistent current fluctuations in equilibrium and under
the harmonic ac pumping.
In Sec.VI we consider the problem of dc conductance under ac pumping 
in two different experimental geometries that correspond to measurements
of the Landauer
and the Kubo conductances.
We re-derive the expression for the Landauer conductance in terms of the
electron Green's function in the time domain and show that the loose
diffusons
cannot be build in this problem. It means that
in systems of non-interacting electrons the Landauer conductance is
insensitive to the electron energy distribution inside the mesoscopic
system and thus is insensitive to heating. In contrast to that the Kubo conductance
is sensitive to the
non-equilibrium electron distribution in the corresponding system. 
In Conclusion we summarize the main results of the paper and point out on
its obvious extensions.

\section{Analytical structure of the nonlinear dynamical response}
In this section we describe the general analytical structure of the
non-linear response {\it of an arbitrary order in the external field} using the
formalism of the analytical continuation
\cite{elias} and the Keldysh diagrammatic technique \cite{keld,RS-1986}.
We will show that causality encoded in the triangular matrix structure of
the Green's functions in the Keldysh technique, allows {\it at most one point}
where the string of retarded electron Green's functions is switched to the
string of advanced Green's functions in the expression for the non-linear response
of an {\it arbitrary order}.
\subsection{Causality of  a non-linear response and the method of
analytical continuation}
The
formalism of analytical continuation is based on the explicit assumption
of causality. 
One starts with the electron Green's function  
${\cal G}_{\varepsilon,\varepsilon-\omega}$ 
defined on the Matsubara
discrete frequencies $\varepsilon_{n}=\pi T (2n+1)$, $\omega_{n}=2\pi T n$
($T$ is the bath temperature) and expanded in series in the ac
field ${\bf A}_{\omega}$. Any (local in time) observable is 
expressed
through the sum
$T\sum_{\varepsilon}{\cal
G}_{\varepsilon,\varepsilon-\omega}$. The non-linear term
of the $k$-th order in this sum is 
$\sum_{\omega^{(i)}}K(i\omega^{(1)},...i\omega^{(k)})\,{\bf
A}_{\omega^{(1)}}...
{\bf A}_{\omega^{(k)}}$,
where $\omega^{(i)}=2\pi T \,n_{i}$ and:
\widetext
\Lrule
\begin{eqnarray}
\label{ser}
K(i\omega^{(1)},...,i\omega^{(k)})=T\sum_{\varepsilon}
G_{0}(i\varepsilon)G_{0}(i\varepsilon-i\omega^{(1)})
G_{0}(i\varepsilon-i\omega^{(1)}-i\omega^{(2)})...
\,G_{0}(i\varepsilon-i\omega).
\end{eqnarray}
\Rrule   
\narrowtext

In the above equation we omitted the current operators $\hat{{\bf j}}$
or the position operators $\hat{{\bf r}}$ coupled to the vector-potential 
${\bf A}_{\omega}$ or to the electric field ${\bf
E}_{\omega}=i\omega{\bf A}_{\omega}$ in the
electron-field interaction:
\begin{eqnarray}
\label{efi}
{\cal H}_{\rm e-f}&=&\left\{\matrix{ -{\bf\hat{j}}{\bf
A}_{\omega} & {\rm transverse\;\; gauge} \cr
-\hat{{\bf r}}\,{\bf E}_{\omega}  & {\rm longitudinal
\;\;gauge} \cr
}\right.
\end{eqnarray}
and introduced the exact electron Green's
function in the absence of the time-dependent perturbation:
\begin{equation}
\label{Gex}
G_{0}(i\varepsilon)=\sum_{m}\frac{\Psi_{m}({\bf r})\Psi_{m}^{*}({\bf
r'})}{i\varepsilon -\varepsilon_{m}},
\end{equation}
where $\Psi_{m}({\bf r})$ is an exact electron wavefunction in the
mesoscopic system that corresponds to the stationary state with the energy 
$\varepsilon_{m}$.

Causality requires the physical (retarded) response function
$K(\omega_{1},...\omega_{k})$ which depends
on the {\it continuous} frequencies $\omega_{i}$, to have no singularities
in the upper half-plain of ${\it each}$ complex variable $\omega_{i}$.  
Thus $K(\omega_{1},...\omega_{k})$ is obtained by the {\it analytical
continuation} of 
$K(i\omega^{(1)},...,i\omega^{(k)})$ from the imaginary discrete points
$\omega_{i}=i\omega^{(i)}$ into the upper half-plane $\Im \omega_{i} >0$. 
In order to implement the analytical continuation one represents the sum over
$\varepsilon_{n}$ in Eq.(\ref{ser}) as a contour integral 
over the contour ${\cal C}$ that comprises all the points
$i\varepsilon_{n}=i\pi T (2n+1)$ (see Fig.1a):
\begin{eqnarray}
\label{serin}
&&K(i\omega^{(1)},...,i\omega^{(k)})=\int_{{\cal
C}}\frac{d\varepsilon}{4\pi
i}\,\tanh\left(\frac{\varepsilon}{2T}
\right)\\ \nonumber
&\times& G_{0}(\varepsilon)G_{0}(\varepsilon-i\omega^{(1)})
G_{0}(\varepsilon-i\omega^{(1)}-i\omega^{(2)})...
\end{eqnarray} 
The next step is to deform the contur along the cuts $\Im\varepsilon=0$,
$\Im
\varepsilon=\omega^{(1)}$, $\Im 
\varepsilon=\omega^{(1)}+\omega^{(2)}$ and so on (Fig.1b).

In doing that one has to take care of the analytical properties of the
electron Green's functions $G_{0}(\varepsilon)$: the integrand in
Eq.(\ref{serin}) should be regular in each strip between the neighbour
cuts. This means that for $\Im\varepsilon$ larger than that of the upmost
cut all the Green's functions $G_{0}$ should be chosen retarded
$G_{0}^{R}$. In the strip just below this cut one Green's function with
the argument that takes zero value on this cut should be switched to the
advanced $G_{0}^{A}$. In the next stripe another retarded Green's
function is switched from the retarded to the advanced one and so on. Thus
for $\Im\varepsilon <0$
all the retarded Green's functions are replaced by the advanced ones.
\begin{figure}[-b]
\centerline{
\epsfig{figure=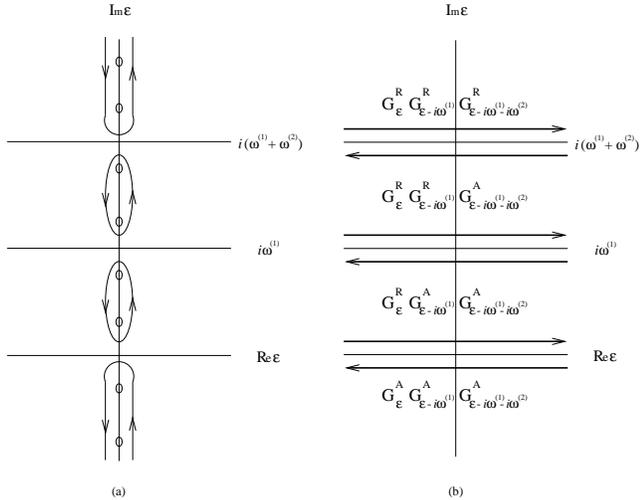,width=20pc,angle=0}
\vspace{5mm}}
\caption{Contours of integration: (a) initial, (b) after deformation
along the cuts. The analytical properties [retarded (R) or advanced (A)]
of Green's functions are also shown in each strip.}
\end{figure}

The next step is to shift the integration from along the cuts to the real
axis
$-\infty< E <+\infty$. To this end we shift the variable of
integration 
$\varepsilon\rightarrow E+i(\omega^{(1)}...+\omega^{(p)})$
corresponding to the $(p+1)$-th cut
$\Im\varepsilon=\omega^{(1)}+...+\omega^{(p)}$. 
Note that since $i\omega^{(i)}$ is a period of $\tanh(\varepsilon/2T)$ the
above shift of variables does not change
$\tanh(\varepsilon/2T)\rightarrow \tanh(E/2T)$. As a
result we get in the r.h.s. of
Eq.(\ref{serin}) a sum of the form:
\begin{eqnarray}
\label{serinre}
\sum_{p=0}^{k}&& \left\{G_{0}^{R}(E+i(\omega^{(1)}+...+\omega^{(p)}))...
G_{0}^{R}(E+i\omega^{(p)})\right.\\ \nonumber
&\times&
[\,G_{0}^{R}(E)-G_{0}^{A}(E)]\,\tanh\left(\frac{E}{2T}
\right)\\
\nonumber
&\times& \left.G_{0}^{A}(E-i\omega^{(p+1)})...
G_{0}^{A}(E-i(\omega^{(p+1)}+...\omega^{(k)}))\right\}.
\end{eqnarray}
Now we are in a position to implement the analytical continuation. It reduces
simply to the replacement $i\omega^{(i)}\rightarrow \omega_{i}$ because
all the
Green's functions have the analytical properties that guarantee the
regularity of each term in the sum Eq.(\ref{serinre}) in the upper
half-plane of each of the complex variables $\omega_{i}$.

One immediately notices the characteristic feature of Eq.(\ref{serinre}):
it is a sum of products (strings) of Green's functions
such that each string is a product of $p$ functions $G_{0}^{R}$
followed by the product of $k-p+1$ functions $G_{0}^{A}$.  
{\it There is only one point in each string where the change of the
character of analyticity
occurs}. This property can be traced back to the causal nature of the
non-linear response.
\subsection{The Keldysh diagrammatic technique}
The analytical structure identical to Eq.(\ref{serinre}) arises in the
Keldysh
technique from the {\it triangular} matrix structure of electron
Green's
functions \cite{LO-1986,RS-1986}:
\begin{eqnarray}
\label{Green-Keldysh}
{\underline G}=
\left(
\begin{array}{cc}   
G^{\rm R} & G^{\rm K} \\
0 & G^{\rm A}
\end{array}
\right),
\end{eqnarray}
where the superscript $K$ stands for the Keldysh component that determines
all the observables $O$:
\begin{eqnarray}
\label{I-definition}
O(t) = i{\rm Tr} \left\{
{\hat O}\,G^{K}(t,t)
\right\}.
\end{eqnarray}
Treating the
applied ac field as a {\it classical} field, we assign
\cite{LO-1986,RS-1986}   
the matrix
vertex $\tau^0 {\cal H}_{\rm e-f}$ to the electron--field interaction
${\cal H}_{\rm e-f}$, where $\tau^0$ is the unit matrix in the 
$2\times 2$ Keldysh
space. 

Using the usual expansion of electron Green's function $\underline{G}$ in
powers of electron--field interaction:
\begin{eqnarray}
\label{I-expansion}
& & O(t) = i\sum_{p=0}^{\infty} {\rm Tr}\left\{{\hat O} \left[
\left\{
{\underline G}_0 {\cal H}_{\rm e-f}
\right\}^p {\underline G}_0
\right]^{\rm K}\right\}_{tt} \\
&=& i\sum_{p=0}^{\infty} \sum_{l=0}^p {\rm Tr}\left[{\hat O} 
\left\{
G_0^{\rm R} {\cal H}_{\rm e-f}
\right\}^l
G_0^{\rm K}
\left\{
{\cal H}_{\rm e-f} G_0^{\rm A}
\right\}^{p-l}
\right]_{tt}, \nonumber
\end{eqnarray}
and the ansatz \cite{LO-1986,RS-1986}:
\begin{equation}
\label{GK}
G_{0}^{K}(E)=\left[G_{0}^{R}(E)-G_{0}^{A}(E)
\right]\,\tanh\left(\frac{E}{2T}
\right)
\end{equation}
one immediately obtains the same R-A structure as in Eq.(\ref{serinre}).

\section{Non-linear response in the time domain}
\subsection{Non-locality of perturbation series in the time domain}

Let us consider the structure of the expansion Eq. (\ref{I-expansion}),
for a given number $p$ of the field vertices, in more
detail. This will later help us to establish the structure of an essentially
non-linear  expessions for different observables in the presence of a time-dependent
ac field.

Three different contributions can be distinguished there. In the
time--domain representation, the first
contribution can be depicted by the electron loop [Fig. 2(a)]. 
\begin{figure}[-b]
\centerline{
\epsfig{figure=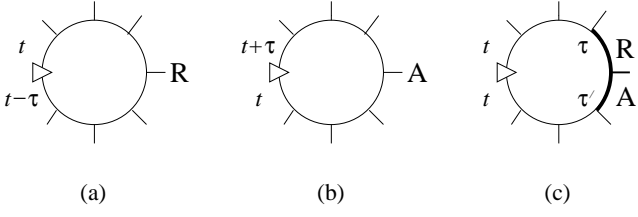,width=20pc,angle=0}
\vspace{5mm}}
\caption{
Graphic representation of the expansion Eq. (\ref{I-expansion}): (a) retarded,
(b) advanced, and (c) retarded--advanced loops.
Rays emanating from the electron loops correspond to the
electron--field interaction ${\cal H}_{\rm e-f}$, and a triangle stands
for the observable operator $\hat{O}$.
In the time domain, the factor ${\hat f}
(\tau)$
is assigned to the $\hat{O}$-vertex in the diagrams (a) and (b).
In the diagram (c) this factor is assigned to the retarded--advanced
junction shown in bold.}
\label{fig2}
\end{figure}

It
entirely consists
of retarded Green's functions:
\widetext
\begin{eqnarray}
\label{R-loop}
+ i \int d\tau {\hat f}(\tau) \int dt_1 \ldots dt_p
{\rm Tr} \left[  
{\hat O}\, G_0^{\rm R}(t,t_1) {\cal H}_{\em e-f}(t_1)
G_0^{\rm R}(t_1,t_2) {\cal H}_{\rm e-f}(t_2) \ldots
G_0^{\rm R}(t_{p-1},t_p) {\cal H}_{\rm e-f}(t_p)
G_0^{\rm R}(t_p, t-\tau)
\right]. 
\end{eqnarray}
The second contribution is associated with the electron loop [Fig. 2(b)]
that solely
contains advanced Green functions:
\begin{eqnarray}
\label{A-loop} 
- i \int d\tau {\hat f}(\tau) \int dt_1 \ldots dt_p
{\rm Tr} \left[
{\hat O}\, G_0^{\rm A}(t+\tau, t_1) {\cal H}_{\rm e-f}(t_1)
G_0^{\rm A}(t_1,t_2) {\cal H}_{\rm e-f}(t_2) \ldots
G_0^{\rm A}(t_{p-1},t_p) {\cal H}_{\rm e-f}(t_p)
G_0^{\rm A}(t_p,t)
\right].
\end{eqnarray}
The third contribution is associated with the electron loop [Fig. 2(c)]
built of
$l$ retarded and $(p-l+1)$ advanced Green functions:
\begin{eqnarray}
\label{RA-loop}
&+& i
\int dt_1 \ldots dt_{l-1}
dt_{l+1} \ldots dt_p
{\rm Tr} \left[
{\hat O}\, G_0^{\rm R}(t,t_1){\cal H}_{\rm e-f}(t_1) \ldots
G_0^{\rm R}(t_{l-2},t_{l-1}){\cal H}_{\rm e-f}(t_{l-1})
\right. \nonumber \\
&\times& \,
\underbrace{\int d\tau d\tau^\prime {\hat f}(\tau-\tau^\prime)\,
G_0^{\rm R}(t_{l-1}, \tau) [{\cal H}_{\rm e-f}(\tau)-{\cal H}_{\rm e-f}(
\tau^\prime)]
G_0^{\rm A}(\tau^\prime, t_{l+1}) }_{\rm retarded-advanced \,\,junction}
\left.
{\cal H}_{\rm e-f}(t_{l+1}) G_0^{\rm A}(t_{l+1}, t_{l+2}) \ldots
{\cal H}_{\rm e-f}(t_p) G_0^{\rm A}(t_p, t)
\right]
\end{eqnarray}
\Rrule
\narrowtext
\noindent  
where
\begin{eqnarray}
\label{f-fourier}
{\hat f}(\tau) = \int \frac{dE}{2\pi} e^{iE\tau} f(E) =
\frac{iT}{\sinh(\pi\tau T)}
\end{eqnarray}
denotes the Fourier transform of $f(E)=\tanh(E/2T)$.

The characteristic feature of the diagrammatic expansion for an observable  
in the time
domain is that there is one special point (ray) on the loop of Fig.2 which
{\it does not} correspond to a single point in the time domain. It is the
point where
the
Fourier
transform of the energy distribution function is assigned to. 
Of special importance
is the {\it retarded-advanced junction}:
\begin{equation}
\label{RAj}
G_{0}^{R}(t,\tau)\,\hat{f}(\tau-\tau')\,[{\cal H}_{\rm e-f}(\tau)-{\cal
H}_{\rm e-f}(\tau')]\,G_{0}^{A}(\tau',t').
\end{equation}
It reveals the non-local in the time domain structure of the point 
[see Fig.2c] where an (arbitrary long) sequence of retarded Green's functions is
switched
to an (arbitrary long)
sequence of advanced Green's functions. 
\subsection{Diffusons and cooperons in the time domain}
With the aim to describe the essentially
non-linear dependence of observables on the external time-dependent field we
introduce the {\it infinite} sequence of retarded (advanced) Green's functions
$G_{0}^{R,A}$:
\begin{equation}
\label{ra}
{\bf G}^{R,A}({\bf r},{\bf
r'};t,t')=\sum_{p=0}^{\infty}\left\{\left[G_{0}^{R,A}\,{\cal
H}_{\rm e-f}\right]^{p}\,G_{0}^{R,A}\right\}_{{\bf r},{\bf r'};t,t'},
\end{equation}
where multiplication assumes the convolution over the coordinate and time
variables $\{AB\}_{{\bf r},{\bf r'};t,t'}=\int d{\bf r''}dt''
A_{{\bf r},{\bf r''}}(t,t'')\,B_{{\bf r''},{\bf r'}}(t'',t')$. 

In describing weak localization and mesoscopic phenomena a special
role is played by the disorder averages of a pair of electron Green's
functions called `diffusons':
\begin{eqnarray}
\label{D}
&&{\cal D}_{{\bf r}{\bf
r'}}(t_{+},t'_{+};t_{-},t'_{-})=\delta(\eta-\eta')\,
D_{\eta}(t,t';{\bf r},{\bf r'})\\ \nonumber
&=& (2\pi\nu\tau_{\rm e})^{-2}\,\langle {\bf G}^{R}({\bf r},{\bf
r'};t_{+},t^{\prime}_{+})   
{\bf G}^{A}({\bf r'},{\bf r};t_{-}^{\prime},t_{-})
\rangle,
\end{eqnarray}
and `cooperons':
\begin{eqnarray}
\label{C}
&&{\cal C}_{{\bf r}{\bf
r'}}(t_{+},t'_{+};t_{-},t'_{-})=\frac{1}{2}\delta(t-t')\,
C_{t}(\eta,\eta';{\bf r},{\bf r'})\\ \nonumber
&=& (2\pi\nu\tau_{\rm e})^{-2}\langle {\bf G}^{R}({\bf r},{\bf
r'};t_{+},t^{\prime}_{+})
{\bf G}^{A}({\bf r},{\bf r'};t_{-},t_{-}^{\prime})
\rangle.
\end{eqnarray}
In Eqs.(\ref{D}, \ref{C}) $\langle ...\rangle$ stands for the disorder
average, $\nu$ is the mean electron density of states, $\tau_{\rm e}$ is
the electron momentum relaxation time, $t_{\pm}=t\pm\eta/2$,
$t'_{\pm}=t'\pm\eta'/2$. The $\delta$-functions in Eqs.(\ref{D}, \ref{C})
result from the assumption on the constant mean density of states over
the relevant energy interval much smaller than the electron bandwidth. 

In the transverse gauge, the functions $D_{\eta}(t,t';{\bf r},{\bf r'})$
and
$C_{t}(\eta,\eta';{\bf r},{\bf r'})$ obey the following equations
\cite{AAK-1982} which correspond to the diffusion approximation with $D$
being the diffusion constant:
\begin{eqnarray}
\label{diff}
\left\{\frac \partial {\partial
t}+\gamma+D\left[i\nabla +
{\bf A}_{{\bf r}}\left(t+
\frac \eta 2\right)-{\bf A}_{{\bf r}}\left(t-\frac \eta 2\right)\right]
^2\right\} \\ \nonumber\times
D_{\eta}(t,t^{\prime };{\bf r},{\bf r'})= \delta (t-t^{\prime
})\frac{\delta({\bf r}-{\bf r'})}{2\pi\nu\tau_{\rm e}^2}.
\end{eqnarray}
and
\begin{eqnarray}\label{coop}
\left\{2 \frac \partial {\partial \eta }+\gamma+ D
\left[i\nabla + {\bf A}_{{\bf r}}\left(t+\frac \eta 2\right)+
{\bf A}_{{\bf r}}\left(t-\frac \eta
2\right)\right]^2\right\}\\ \nonumber\times
C_{t}(\eta, \eta ^{\prime };{\bf r},{\bf r'})=2 \delta (\eta -\eta
^{\prime
})\frac{\delta({\bf r}-{\bf r'})}{2\pi\nu\tau_{\rm e}^2},  
\end{eqnarray}
where we assume the electron-field interaction  Eq.(\ref{efi})
corresponding to the transverse gauge
with the weak space (on the scale of the elastic mean free path
$\ell=v_{F}\tau_{\rm e}$)
and time (on the scale of $\tau_{\rm e}$) dependence of the external
classical field ${\bf A}_{{\bf r}}(t)$ which is also supposed to be weak
enough $|{\bf A}_{{\bf r}}(t)|\ell\ll 1$.
We also assume the possibility for
electrons to escape the mesoscopic system which is described by the
(small) escape rate $\gamma$. 
\subsection{The ring and the quantum dot
geometry} 
Equations (\ref{diff}, \ref{coop}) should be supplemented by the boundary
conditions. Below we consider two principally different geometries 
shown schematically in Fig.3.
\begin{figure}[-b]
\centerline{
\epsfig{figure=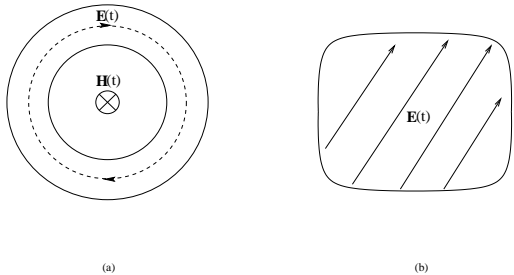,width=16pc,angle=0}
\vspace{5mm}
}
\caption{The ring (a) and the quantum dot (b) geometry} 
\end{figure}
One of them corresponds to the mesoscopic ring of the circumference $L$
with a
small aspect ratio pierced by
the time-dependent magnetic field ${\bf H}(t)$ which generates the
circular  
electric field ${\bf E}(t)= - \partial {\bf A}/\partial t$, where
${\bf A}(t)= \frac{1}{2} {\bf H}\times {\bf r}$. 
In this {\it ring geometry} the diffusons and cooperons should obey the
{\it periodic} boundary conditions.

The ring
geometry
corresponds to the pumping by the {\it magnetic part} of the microwave
field 
when the size of the mesoscopic system $L$ is less than the size of the
skin layer. 
However, for a semiconductor quantum dot with not too large
conductivity $\sigma$ the {\it electric part} of the microwave field can
also
penetrate inside the mesoscopic system. 
For the usual situation $L\ll
\lambda$ where $L$ is the size of the system and $\lambda$ is the 
microwave wavelength the electric field inside the system ${\bf
E}(t)=-\partial {\bf A}/\partial t$ is
homogeneous in space.

In this situation the boundary conditions read:
\begin{eqnarray}
\label{BCC}
&&\left[i\nabla + {\bf Q}^{(d)}(t,\eta)\right]_{{\bf n}}\,
D_{\eta}(t,t^{\prime };{\bf r},{\bf r'})=0,
\\ \nonumber 
&& \left[i\nabla + {\bf Q}^{(c)}(t,\eta)\right]_{{\bf
n}}\,C_{t}(\eta, \eta ^{\prime };{\bf r},{\bf
r'})=0,
\end{eqnarray} 
where ${\bf n}$ is the vector normal to the boundary at a point ${\bf
r}$ and we introduce a short-hand notation:
\begin{equation}
\label{Qd}
{\bf Q}^{(d)}(t,\eta)={\bf A}\left(t+
\frac \eta 2\right)-{\bf A}\left(t-\frac \eta 2\right),
\end{equation}
\begin{equation}
\label{Qc}
{\bf Q}^{(c)}(t,\eta)={\bf A}\left(t+
\frac \eta 2\right)+{\bf A}\left(t-\frac \eta 2\right).
\end{equation}
\subsection{Diffusons and cooperons in the ring geometry}
In this case the convenient coordinate $x$ along the ring is proportional
to the azimuthal angle $x=L\theta/2\pi$. One can suppress the transverse
coordinate, as the dependence on that coordinate at a small aspect ratio
is negligible. The tangential component of the field $A_{x}$
is independent on $x$. Given the periodic boundary conditions
for diffusons and cooperons, one can switch to the Fourier-transforms 
$D_{\eta}(t,t^{\prime };q)$ and $C_{t}(\eta,\eta^{\prime };q)$ in
Eqs.(\ref{diff}, \ref{coop}) with the quantized momentum
$q_{m}=(2\pi/L)\,m$, where $m=0,\pm1,\pm2...$.
Then the solution is straightforward:
\begin{eqnarray}
\label{PBCs}
&&D_{\eta}(t,t^{\prime
};q)=\frac{\theta_{t-t'}\,e^{-\gamma (t-t')}}{2\pi\nu\tau_{\rm
e}^2}\,e^{-D\int_{t'}^{t}d\tau\,\left[q-Q^{(d)}(\tau,\eta)
\right]^2},\\ \nonumber
&&C_{t}(\eta,\eta^{\prime
};q)=\frac{\theta_{\eta-\eta'}\,e^{-\frac{1}{2}\gamma
(\eta-\eta')}}{2\pi\nu\tau_{\rm
e}^2}\,e^{-\frac{D}{2}\int_{\eta'}^{\eta}d\tau\,
\left[q-Q^{(c)}(t,\tau)
\right]^2},
\end{eqnarray}
where $\theta_{t}$ is the step function and $Q^{(d,c)}$ is defined by
Eqs.(\ref{Qd}, \ref{Qc}). 

An important particular case is the {\it zero-mode} diffusons
$D_{\eta}(t,t^{\prime})=D_{\eta}(t,t^{\prime};q=0)$ and
cooperons $C_{t}(\eta,\eta^{\prime})=C_{t}(\eta,\eta^{\prime};q=0)$ which
correspond to $q=0$:
\begin{eqnarray}
\label{0d}
&&D_{\eta}(t,t^{\prime})=\frac{\theta_{t-t'}e^{-\gamma
(t-t')}}{2\pi\nu\tau_{\rm
e}^2}\,e^{-D\int_{t'}^{t}d\tau\,\left[{\bf Q}^{(d)}(\tau,\eta)
\right]^2},\\ \nonumber
&&
C_{t}(\eta,\eta^{\prime})=\frac{\theta_{\eta-\eta'}e^{-\frac{1}{2}\gamma(\eta-\eta')}}{2\pi\nu\tau_{\rm
e}^2}\,e^{-\frac{D}{2}\int_{\eta'}^{\eta}d\tau\,
\left[{\bf Q}^{(c)}(t,\tau)
\right]^2}
\end{eqnarray}
One can see
that they decay as a
function of $t-t'$ or $\eta-\eta'$ even at an escape rate $\gamma=0$.
This is the manifestation of {\it dephasing} by the time-dependent
external field. We note that Eq.(\ref{D}) with $\eta=0$ corresponds to
the electron density correlation function. In the absence of electron
escape the
total number of particles is conserved and
therefore $D_{\eta=0}(t,t^{\prime};q=0)$ must be a constant for any
$t>t'$. This is consistent with the property of $Q^{(d)}(\tau,\eta=0)=0$.
However, there is no constraint that would prohibit a decay of 
$D_{\eta}(t,t^{\prime})$ at a non-zero $\eta$. 
\subsection{Diffusons and cooperons in the quantum dot geometry.}
This case is principally different because of the field-dependent boundary
conditions Eqs.(\ref{BCC}) and the potential (longitudinal) nature of the 
electric field inside the dot. This makes the description in the
longitudinal gauge more
convenient in the quantum dot geometry. 
\subsubsection{Equations for the diffusons and the cooperons in the
longitudinal gauge}

Performing the  gauge transformation:
\begin{equation}
\label{gtd}
D_{\eta}(t,t^{\prime };{\bf r},{\bf
r'})=e^{i{\bf Q}^{(d)}(t,\eta){\bf r}}\,\tilde{D}_{\eta}(t,t^{\prime
};{\bf r},{\bf r'})\,e^{-i{\bf Q}^{(d)}(t',\eta){\bf r'}}, 
\end{equation}
\begin{equation}
\label{gtc}
C_{t}(\eta,\eta^{\prime };{\bf r},{\bf
r'})=e^{i{\bf Q}^{(c)}(t,\eta){\bf r}}\,\tilde{C}_{t}(\eta,\eta^{\prime
};{\bf r},{\bf r'})\,e^{-i{\bf Q}^{(c)}(t,\eta'){\bf r'}},
\end{equation}
we switch to the longitudinal gauge.

This transformation removes the time-dependent field from the boundary
conditions Eq.(\ref{BCC}). However, it remains in the equations for
$\tilde{D}_{\eta}(t,t^{\prime
};{\bf r},{\bf r'})$ and $\tilde{C}_{t}(\eta,\eta^{\prime
};{\bf r},{\bf r'})$ but only as the time-derivatives:
\begin{eqnarray}
\label{dot}
\dot{{\bf Q}}(t,\eta)&=&\frac{\partial}{\partial t}\, {\bf
Q}^{(d)}(t,\eta)=2\frac{\partial}{\partial \eta}\,
{\bf Q}^{(c)}(t,\eta)\\ \nonumber
&=& {\bf E}\left(t-\frac{\eta}{2}\right)-{\bf
E}\left(t+\frac{\eta}{2}\right),
\end{eqnarray}
where ${\bf E}(t)=-\partial {\bf A}(t)/\partial t $.

The operators in the l.h.s.
of the corresponding
equations take the form:
\begin{eqnarray}
\label{longD} 
\hat{{\cal L}}_{d}=\left\{\frac \partial {\partial
t}+\gamma-D\nabla^2 + i{\bf r}\dot{{\bf Q}}(t,\eta)
\right\}
\end{eqnarray}
\begin{eqnarray}
\label{longC}
\hat{{\cal L}}_{c}=\left\{2\frac \partial {\partial
\eta}+\gamma-D\nabla^2 + i{\bf r}\dot{{\bf Q}}(t,\eta)
\right\}.
\end{eqnarray}
It is convenient to expand  $\tilde{D}_{\eta}(t,t';{\bf r},{\bf r'})$ and
$\tilde{C}_{t}(\eta,\eta';{\bf r},{\bf r'})$ in an infinite sum
$\sum_{\nu,\mu}A_{\nu\mu}\,\Phi_{\nu}({\bf r})\Phi_{\mu}(\bf r')$ over
eigenfunctions $\Phi_{\nu}({\bf r})$ of the diffusion operator
$-D\nabla^2$ with the
Neumann boundary
condition [Eq.(\ref{BCC}) with ${\bf Q}^{(d,c)}=0$]. Then the equation for
the amplitudes $A_{\nu\mu}$ is of the form (for simplicity we consider the
case $\gamma=0$ and suppress the redundant indices and variables):
\begin{equation}
\label{A}
\frac{\partial}{\partial t}\,A_{\nu\mu}+E_{\nu}\,A_{\nu\mu}+i\dot{{\bf
Q}}(t)\,\sum_{\lambda}{\bf
r}_{\nu\lambda}A_{\lambda\mu}=\delta(t-t')\delta_{\nu\mu},
\end{equation}
where ${\bf r}_{\nu\lambda}$ is the matrix element of the vector ${\bf r}$
in the
basis of $\Phi_{\nu}({\bf r})$ and $E_{\nu}$ is the eigenvalue that
corresponds to the eigenfunction $\Phi_{\nu}({\bf r})$.

Though this equation looks similar to the (imaginary time) Schr\"odinger
equation for the {\it dynamical Stark
effect}, identifying $\dot{{\bf Q}}$
as the actual electric field inside the dot would lead to a mistake.  
Let us consider an important case of the {\it constant in time} electric
field ${\bf E}$ which corresponds to ${\bf A}(t)=-{\bf E}t$. Then
Eqs.(\ref{Qd}, \ref{Qc}) give ${\bf Q}^{(d)}(t,\eta)=-{\bf E}\eta$ and
${\bf Q}^{(c)}(t,\eta)=-2{\bf E}t$. One can see from Eq.(\ref{dot}) that
in both cases the
corresponding time-derivatives $\dot{{\bf Q}}^{(d)}$ and $\dot{{\bf
Q}}^{(c)}$ are {\it identically zero}!

The conclusion which can be immediately drawn from this observation is
that the {\em constant in time longitudinal (potential) electric field
cannot lead
to a dephasing} [cf. Ref.\cite{cast}]. This statement is not true for the
circular electric
field
considered above. This field is not constant in the Cartesian coordinates  
and is not potential ${\bf curl}\;\;{\bf E}\neq 0$.
\subsubsection{The weak-field adiabatic and anti-adiabatic limits}
The general solution to Eq.(\ref{A})is unknown even for the
space-homogeneous electric field ${\bf E}(t)$ and in the ergodic limit
\begin{equation}
\label{erg}
E_{c}(t-t')\gg 1\;\;\;\; E_{c}(\eta-\eta')\gg 1. 
\end{equation}
However, one can find a
simple approximate solutions in the weak field limit
\begin{equation}
\label{wfl}
\frac{|\dot{{\bf Q}}|\,L}{E_{c}}\ll 1.
\end{equation}
In the ergodic limit Eq.(\ref{erg}) the diffusons
$\tilde{D}_{\eta}(t,t';{\bf r},{\bf r'})$ and cooperons
$\tilde{C}_{t}(\eta,\eta';{\bf r},{\bf r'})$ are nearly
space-independent, i.e. the corresponding expansions are dominated by the
{it zero-mode} $\Phi_{0}$ with $E_{0}=0$. By definition the next mode has
the
eigenvalue equal to the Thouless energy $E_{1}=E_{c}$. 
For the corresponding amplitude $A_{00}$ we get from Eq.(\ref{A}):
\begin{equation}
\label{00}
\frac{\partial}{\partial t}A_{00}
+
i\dot{{\bf Q}}(t)\sum_{\mu\neq 0}{\bf r}_{0\mu}\,A_{\mu 0}=\delta(t-t'),
\end{equation}
where we choose the system of coordinates in which ${\bf r}_{00}=0$.

In the weak field limit Eq.(\ref{wfl}) one can neglect in Eq.(\ref{A})
$\dot{{\bf
Q}}{\bf r}_{\mu\nu}A_{\nu 0}$ compared to $E_{\mu}A_{\mu 0}$
($\mu,\nu\neq 0$) and obtain the closed system of equations for $A_{00}$
and $A_{\mu 0}$:
\begin{equation}
\label{mu0}
\frac{\partial}{\partial t}A_{\mu 0}+E_{\mu} A_{\mu 0}+i\dot{{\bf Q}}(t)
{\bf r}_{\mu 0}\,A_{00}=0.
\end{equation} 
Solving Eq.(\ref{mu0}) and substituting $A_{\mu 0}$ into Eq.(\ref{00}) 
we obtain:
\widetext
\begin{eqnarray}
\label{wfeq}
\frac{\partial}{\partial t}A_{00}
+\int_{t'}^{t}\dot{{\bf
Q}}(t){\bf\Delta}(t-t'')\dot{{\bf
Q}}(t'')\,A_{00}(t'',t')\,dt''=\delta(t-t'),
\end{eqnarray}
\Rrule
\narrowtext
where ${\bf\Delta}(t)$ is the matrix in the vector space:
\begin{equation}
\label{del}
[{\bf\Delta}(t)]_{ij}=\sum_{\mu\neq 0}[{\bf
r}_{0\mu}]_{i}\,e^{-E_{\mu}t}\,[{\bf r}_{\mu 0}]_{j}.
\end{equation}
Eq.(\ref{wfeq}) can be further simplified in the {\it adiabatic} limit
where $\dot{{\bf Q}}(t)$ is dominated by the frequencies $\omega\ll
E_{c}$. In this case the quantity Eq.(\ref{del}) can be approximated by
the $\delta$-function $[{\bf\Delta}(t)]_{ij}=C_{ij}\delta(t)$, where:
\begin{equation}
\label{Cij}
C_{ij}=\sum_{\mu\neq 0} [{\bf 
r}_{0\mu}]_{i}\,[E_{\mu}]^{-1}\,[{\bf r}_{\mu 0}]_{j}.
\end{equation}
Then Eq.(\ref{wfeq}) can be immediately solved and we obtain for the
zero-mode amplitudes:
\widetext
\Lrule
\begin{eqnarray}
\label{difad}
A^{(d)}_{00}(t,t';\eta)=\tilde{D}_{\eta}(t,t^{\prime
})=\frac{\theta_{t-t'}}{2\pi\nu\tau_{\rm
e}^2}\,
\exp\left[-\int_{t'}^{t}\dot{{\bf Q}}^{(d)}_{i}(t'',\eta)\,C_{ij}
\dot{{\bf Q}}^{(d)}_{j}(t'',\eta)\,dt''
\right]. 
\end{eqnarray}
\begin{eqnarray}
\label{coopad}
A^{(c)}_{00}(\eta,\eta';t)=\tilde{C}_{t}(\eta,\eta^{\prime
})=\frac{\theta_{\eta-\eta'}}{2\pi\nu\tau_{\rm e}^2}\,
\exp\left[-\frac{1}{2}\int_{\eta'}^{\eta}\dot{{\bf
Q}}^{(c)}_{i}(t,\eta'')\,C_{ij}
\dot{{\bf Q}}^{(c)}_{j}(t,\eta'')\,d\eta''
\right]. 
\end{eqnarray}
\Rrule
\narrowtext
Eqs.(\ref{difad}),(\ref{coopad}) are valid in the adiabatic limit
$\omega\ll E_{c}$ provided that the conditions 
Eqs.(\ref{erg}),(\ref{wfl}) are fulfilled.

In the opposite {\it anti-adiabatic} limit $\omega\gg E_{c}$, the
integral term
in Eq.(\ref{wfeq}) is recast as the total time-derivative (which has zero
time average and thus strongly oscillates) and the remainder consisting of
the strongly oscillating term $-{\bf Q}(t){\bf \Delta}(0)\dot{\bf
Q}(t)A_{00}(t,t')$ and the term  
$$
-\int_{t'}^{t}dt''\;{\bf Q}(t)\,[\partial/\partial t {\bf
\Delta}(t-t'')]\,\dot{{\bf
Q}}(t'')A_{00}(t'',t').
$$
that contains a weakly oscillating part. Integrating this term by parts
and using the identity $\dot{{\bf \Delta}}(0)=\partial/\partial t [{\bf
\Delta}]_{ij}|_{t=0}=-D\delta_{ij}$ one extracts this weakly
oscillating part:
\begin{equation}
\label{wop}
-{\bf Q}(t)\dot{{\bf \Delta}}(0)\,{\bf Q}(t)\, A_{00}(t,t')=D\,[{\bf
Q}(t)]^{2}\,A_{00}(t,t').
\end{equation}
Not surprisingly, we conclude that in the anti-adiabatic weak-field limit
the system does not feel the boundary and the zero-mode
diffusons and cooperons in the quantum dot geometry coincide with those
in the ring geometry Eqs.(\ref{0d}). 

However, in the adiabatic weak-field limit these two geometries are
principally different, since the dephasing
factor in Eq.(\ref{difad}),(\ref{coopad}) contains a quadratic form in
the time-derivative $\dot{{\bf Q}}$ and a structural constant that depends
on the system size $L$, while the dephasing 
factor in Eqs.(\ref{0d}) contains a quadratic form in ${\bf Q}$ and is
independent of $L$.
\subsection{Time-dependent random matrix theory (TRMT)}

The zero-mode approximation Eq.(\ref{erg}) is equivalent
to the {\it random-matrix theory} (RMT). The fact that there are ${\it
two}$
different forms, Eqs.(\ref{difad}),(\ref{coopad}) and Eqs.(\ref{0d}), of diffusons
and cooperons in the zero-mode approximation
suggests two different ways of defining the {\it time-dependent} RMT.
The idea is to define the Gaussian ensembles of {\it time-dependent} random matrices 
which reproduce the expressions Eq.(\ref{difad}),(\ref{coopad}) or Eqs.(\ref{0d}).
Then so defined TRMT can be applied to describe not only disodered mesoscopic
systems with the diffusion motion of electrons (used in the above derivation) but
also ballistic quantum dots with the chaotic electron motion.

Before proceeding with the formal derivation we address a possible confusion based
on the common wisdom that only systems in the external field with the characteristic
frequency $\omega\ll
E_{c}$ can be described by the random matrix theory. This statement is valid for
a linear response but it is incorrect in the non-linear case. The point is that in
the linear case the frequency of the external field enters {\it all} diffusons
or
cooperons  as $[Dq^2 -i\omega]^{-1}$ where $\omega$ is the difference
between the energy variables of retarded and advanced electron Green's
functions. Assuming the summation over momenta $q$ and
the fact that the first non-zero mode corresponds to $Dq^2 =E_{c}$ one concludes
that
at  $\omega\ll E_{c}$ the main contribution to the sum over momenta is given by the
zero mode with $q=0$. In the non-linear case the situation is more complicated,
since ${\cal H}_{\rm e-f}(t)$ in Eq.(\ref{ra}) is a sum of two parts
proportional to $e^{i\omega t}$ and
$e^{-i\omega t}$. As a result of the frequency fusion $\omega-\omega=0$ in the
field-dependent
diffuson(cooperon) self-energy part the difference between the energy variables
of retarded and advanced electron Green's
functions constituting a difuson(cooperon) may be zero despite $\omega\gg E_{c}$.
This is the reason why the high-frequency external field modifies the zero-mode
approximation but does not kill it. The two modifications of the TRMT are:

(i) For the conductors of the {\it ring topology} and the {\it quantum dots in 
the anti--adiabatic limit}, $\omega \gg E_{c}$, the time dependent RMT is 
defined by the matrix Hamiltonian 
\begin{equation}
\label{RMT2}
H=H_{0}+i\xi(t)V_{a},
\end{equation} 
where $\xi(t)$ is a real--time dependent function, 
$N\times N$ random matrix $H_{0}$ belongs to the Gaussian orthogonal 
ensemble while $V_{a}$ is the real {\it anti-symmetric} random matrix 
(which corresponds to the sign $-$ in Eq.(\ref{mdef})) of the same size,
\begin{eqnarray}
    \label{mdef}
    \langle H_{0}^{nm}H_{0}^{n^{\prime}m^{\prime}}\rangle
    = \frac{N\delta^{2}}{\pi^{2}} 
    [\delta_{mm^{\prime}}\delta_{nn^{\prime}} + 
    \delta_{mn^{\prime}}\delta_{nm^{\prime}}], \nonumber \\
    \langle V_{a,s}^{nm}V_{a,s}^{n^{\prime}m^{\prime}}\rangle
    = \frac{\delta C}{\pi} 
    [\delta_{mm^{\prime}}\delta_{nn^{\prime}} \mp 
    \delta_{mn^{\prime}}\delta_{nm^{\prime}}],
\end{eqnarray}
$\delta$ being the mean level spacing, and $C$ is a non-universal 
constant to be identified later on.
To make a link between the TRMT, Eqs. (\ref{RMT2}) and (\ref{mdef}), 
and the zero--dimensional approximation Eqs. (3.15) and (3.31), it is 
useful to introduce the TRMT diffuson, $D_{\eta}(t,t^{\prime})$, and 
cooperon, $C_{t}(\eta,\eta^{\prime})$, propagators
\widetext
\Lrule
\begin{eqnarray}
\label{cd-2}
\sum_{n,m}\langle 
G_{nm}^{R}(t_{+},t_{+}^{\prime})
G_{mn}^{A}(t_{-}^{\prime},t_{-})
\rangle = \left(2\pi \nu \tau_{\rm e}(N)\right)^{2} 
\delta(\eta-\eta^{\prime}) 
D_{\eta}(t,t^{\prime}), \nonumber \\
\sum_{n,m}\langle 
G_{nm}^{R}(t_{+},t_{+}^{\prime}) G_{nm}^{A}(t_{-},t_{-}^{\prime})
\rangle = \frac{1}{2}\left(2\pi \nu \tau_{\rm e}(N)\right)^{2}
\delta(t-t^{\prime}) 
C_{t}(\eta,\eta^{\prime}),
\end{eqnarray}
where $G^{R,A}(t,t')$ are retarded or advanced Green's functions that
correspond to the matrix Hamiltonian H, $t_{\pm}=t\pm\eta/2, 
t_{\pm}^{\prime}=t^{\prime}\pm\eta^{\prime}/2$; $\tau_{\rm 
e}(N)=\pi/(2N\delta)$, and $\nu=1/\delta$.
Using the standard method of Refs. \cite{AAK-1982} (see also
Ref.\cite{VA}), we derive the following 
equations for the TRMT propagators:

\begin{eqnarray}
    \label{d-eq}
    \left\{
    \frac{\partial}{\partial t} + C \left[
    \xi\left(t+\frac{\eta}{2}\right) -\xi
    \left(
    t-\frac{\eta}{2}
    \right)
    \right]^{2}
    \right\}
    D_{\eta}(t,t^{\prime}) = \frac{\delta(t-t^{\prime})}
    {2\pi\nu \tau_{e}^{2}(N)}, \\
    \label{c-eq}
    \left\{
    \frac{\partial}{\partial \eta} + \frac{C}{2} \left[
    \xi\left(t+\frac{\eta}{2}\right) +\xi
    \left(
    t-\frac{\eta}{2}
    \right)
    \right]^{2}
    \right\}
    C_{t}(\eta,\eta^{\prime}) = \frac{\delta(\eta-\eta^{\prime})}
    {2\pi\nu \tau_{\rm e}^{2}(N)}.
\end{eqnarray}
\narrowtext
Comparison with the microscopic Eqs. (3.9) and (3.10) [or (3.31)] 
suggests identifying the phenomenological constant $C$ introduced in Eq. 
(\ref{mdef}) with the diffusion coefficient $D$; the 
time--dependent function $\xi(t)$ plays the role of the vector 
potential $A(t)$. This establishes the equivalence between the TRMT and 
the zero--dimensional limit of the microscopic theory of Sec. III on 
the {\it perturbative} level.

(ii) For the {\it quantum dots in 
the adiabatic limit}, $\omega \ll E_{c}$, the time dependent RMT is 
defined by the matrix Hamiltonian \cite{VA,VAA}
\begin{equation}
\label{RMT1}
H=H_{0}+\xi(t)V_{s},
\end{equation} 
where $\xi(t)$ is a real--time dependent function, $H_{0}$ and $V_{s}$ are 
statistically independent, $N\times N$, random matrices belonging to the 
Gaussian orthogonal 
ensemble [ see Eq. (\ref{mdef}) with the sign $+$].

Again, a link between the TRMT, Eqs. (\ref{RMT1}) and (\ref{mdef}), 
and the zero--dimensional approximation Eqs. (3.29) and (3.33) of the
full microscopic theory, is easily established by deriving the equations 
for TRMT propagators. One obtains [see also Refs.\cite{VA,VAA}]:
\widetext
\Lrule
\begin{eqnarray}
    \label{d-eq-22}
    \left\{
    \frac{\partial}{\partial t} + C \left[
    \xi\left(t+\frac{\eta}{2}\right) -\xi
    \left(
    t-\frac{\eta}{2}
    \right)
    \right]^{2}
    \right\}
    D_{\eta}(t,t^{\prime}) = \frac{\delta(t-t^{\prime})}
    {2\pi\nu \tau_{e}^{2}(N)}, \\
    \label{c-eq-22}
    \left\{
    \frac{\partial}{\partial \eta} + \frac{C}{2} \left[
    \xi\left(t+\frac{\eta}{2}\right) -\xi
    \left(
    t-\frac{\eta}{2}
    \right)
    \right]^{2}
    \right\}
    C_{t}(\eta,\eta^{\prime}) = \frac{\delta(\eta-\eta^{\prime})}
    {2\pi\nu \tau_{\rm e}^{2}(N)}.
\end{eqnarray}
\Rrule
\narrowtext
Comparing with the microscopic Eqs. (3.29) and (3.30), we conclude 
that the non-universal constant $C$ of Eq. (\ref{mdef}) has to be 
identified with the one given by Eq. 
(3.28); the function $\xi(t)$ plays the role of ac electric field $E(t)$
in Eq. (3.18). Thus, equivalence between the microscopic approach of 
Sec. III and the TRMT of the form Eq. (\ref{RMT1}) is proven 
perturbatively.

\section{The `loose' diffusons}
In this section we show that starting from the quadratic in the external field
order the diagrams of the impurity technique acquire a new feature: one can 
draw the diffuson with the free end (`the loose diffuson') which carries zero
momentum and zero frequency. We will show below that it is exactly the element
which describes heating by the external field.
\subsection{The loose diffusons and the retarded-advanced junctions}
The analytical structure of the retarded-advanced junction Eq.(\ref{RAj})
leads after the disorder averaging to an unusual object,
the {\it `loose' diffuson}. Let us consider this object for the simplest
ring geometry.
\subsubsection{Loose diffusons in the ring geometry} 
Consider the part of a diagram for an
observable or a product of observables that
contains the retarded-advanced junction (Fig.4a).
\widetext
\begin{figure}[-b]
\centerline{
\epsfig{figure=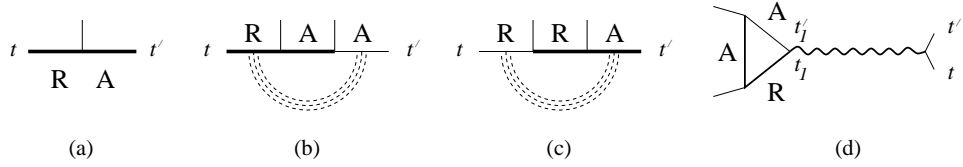,width=30pc,angle=0}
\vspace{5mm}}
\caption{
The retarded--advanced junction (a), and the Green functions'
disorder averaging [(b)
and (c)] resulting in the `loose' diffuson shown by the wavy line (d). It
ends up with the
triangle; 
one of the
triangle
apexes is built upon the retarded--advanced junction, see (b) and (c).
The diagram (d) corresponds to the disorder averaging depicted in (b).
Dashed lines  
in (b) and
(c) denote the diffuson. Because of the vector nature of the vertex  ${\cal
H}_{\rm e-f}$ in the transverse gauge Eq.(\ref{efi}) the loose diffuson should
`embrace'
two vertices  ${\cal 
H}_{\rm e-f}$ in order for the integral over the directions of electron velocity to
be
non-zero. 
}
\label{fig10}
\end{figure}
\narrowtext
\noindent
One can isolate the retarded-advanced junction from the rest of the
diagram by performing the disorder averaging as shown in Fig.4b,c. 
As the result the `loose diffuson' is formed (Fig.4d) 
which originates from the main body of the diagram and terminates at a
triangle
that
consists
of the retarded-advanced junction and another field vertex adjacent to it.
Given the condition $|{\bf A}_{{\bf r}}(t)|\ell\ll 1$ one can neglect the
loose diffusons terminating by a polygon with the number of field vertices
larger than two.

We stress that the loose diffuson can be built only using the
retarded-advanced junction. Indeed, the triangle in Fig.4d whose edges 
correspond to the average retarded and advanced Green functions 
$\bar{G}_{E}^{\rm R(A)}({\bf p})=\int d{\bf r}\langle G_{0}^{\rm
R(A)}({\bf r},{\bf
r}';E)\rangle\,e^{-i{\bf p}({\bf r}-{\bf r}')} 
=(E-\xi_{\bf p}\pm i/(2\tau_{\rm e}))^{-1}$,
describes the electron
motion on the ballistic scale with the electron momentum
relaxation time $\tau_{\rm e}$ being the smallest time scale in the
problem. One may effectively approximate the average Green's functions
in the time-momentum representation $\bar{G}^{\rm R(A)}({\bf p},t)$ by
the $\delta$--functions of the form
$\bar{G}^{\rm R(A)}({\bf p},t) \simeq \bar{G}_{E=0}^{\rm R(A)}({\bf
p})\delta(t)$.
With this approximation and Eq.(\ref{RAj}) the triangle
in Fig.4d reduces to the quadratic in ${\bf A}(t)$ combination
\begin{equation}
\label{ft}
[{\bf A}(t_{1}')-{\bf A}(t_{1})]\,{\bf A}(t_{1}')\,\hat{f}(t_{1}'-t_{1})
\end{equation}
multiplied by a constant ($d$ is the dimensionality of momentum space)
\begin{equation}
\label{trian}
(v_{F}^{2}/d)\nu\int d\xi({\bf p}) \bar{G}_{E=0}^{\rm R}({\bf p})\,
[\bar{G}_{E=0}^{\rm
A}({\bf p})]^2
=2\pi i \nu D \tau_{\rm e}.
\end{equation}
The triangle that corresponds to Fig.4c is given by:
\begin{equation}
\label{trian2}
-2\pi i \nu D\tau_{\rm e}\,[{\bf A}(t_{1}')-{\bf A}(t_{1})]\,{\bf
A}(t_{1})\,\hat{f}(t_{1}'-t_{1}).
\end{equation}

Without the retarded-advanced junction, all the Green's functions in the
triangle would have the same analyticity ($\rm R$ or $\rm A$), and  
the integral over $\xi({\bf p})$ vanishes.

Since in the dynamical approach (e-e and e-ph interactions are neglected) there is
at most one retarded-advanced
junction per electron loop
[see Eq.(\ref{RA-loop})], there could be not more than one loose diffuson
for  a diagram describing the disorder-average of a single observable
$\langle O(t)\rangle$
and not more than $k$ loose diffusons for a diagram
describing the disorder-average of a product of $k$ observables $\langle 
O_{1}(t_{1})...O_{k}(t_{k})\rangle$. 

Each of them is given by:
\begin{eqnarray}
\label{LD}
{\cal D}(t,\eta)&=&2\pi i \nu \tau_{\rm
e} D\,\int dt_{1}\int d{\bf
r}_{1}[{\bf
A}(t_{1}+\eta)-{\bf A}(t_{1})]^2 \nonumber \\
&\times&\hat{f}(\eta)\,D_{-\eta}\left(t+\frac{\eta}{2},
t_{1}+\frac{\eta}{2};{\bf r}_{1}-{\bf r}\right),
\end{eqnarray}
where $\eta=t'-t=t_{1}'-t_{1}$ and  both contributions
Eqs.(\ref{trian}),(\ref{trian2}) to the triangle have been summed up
together. In the ring geometry 
only the zero mode part of $D_{\eta}(t,t';{\bf
r}_{1}-{\bf r})$ survives integration over the coordinate ${\bf r}_{1}$ of
the free end and using Eq.(\ref{0d}) we finally obtain ${\cal
D}(t,\eta)=i\tau_{\rm
e}^{-1}\,\hat{f}(\eta)\Lambda_{\eta}(t)$, where: 
\widetext
\begin{eqnarray}
\label{Lambda-fun}
\Lambda_\eta (t) = D\int_{-\infty}^{t} d\xi\, [{\bf
A}(\xi+\eta)-{\bf A}(\xi)]^2 \, e^{-\gamma (t-\xi)}\,
\exp{\left\{
-D\int_\xi ^{t} d\xi' [{\bf A}(\xi'+\eta)-{\bf A}(\xi')]^2
\right\}}.
\end{eqnarray}
\Rrule
\narrowtext
\noindent
\subsubsection{Loose diffusons in the quantum dot geometry}
Eq.(\ref{Lambda-fun}) holds also in the quantum dot geometry in the {\it
anti-adiabatic} case $\omega\gg E_{c}$. In the opposite
{\it adiabatic} case $\omega\ll E_{c}$ the
longitudinal gauge
is more convenient than the
transverse gauge originally accepted in the paper.
Eqs.(\ref{difad}),(\ref{coopad}) are nothing but the diffusons and
cooperons in the {\it longitudinal gauge} in the zero-mode approximation.
Using Eq.(\ref{RAj}) with ${\cal H}_{\rm e-f}$ corresponding to the
longitudinal gauge Eq.(\ref{efi}) the loose diffuson can be
represented in terms of the matrix elements
$A^{(d)}_{0\mu}$ and $r_{\mu 0}$ as follows:
\begin{eqnarray}
\label{ldif} 
{\cal D}(t,\eta)&=&2\pi i \nu \tau_{\rm e}\hat{f}(\eta)\int
dt_{1}\,[\partial {\bf
A}(t_{1}+\eta)/\partial
t_{1} - \partial {\bf
A}(t_{1})/\partial
t_{1} ]\nonumber \\
&\times&
\sum_{\mu\neq 
0}
A^{(d)}_{0\mu}\left(t+\frac{\eta}{2},t_{1}+\frac{\eta}{2};-\eta\right)\,{\bf
r}_{\mu 0},
\end{eqnarray}
where $A^{(d)}_{0\mu}$ obeys Eq.(\ref{mu0}) with $\dot{{\bf
Q}}(t)=\dot{{\bf Q}}(t,-\eta)$ and $A^{(d)}_{00}$ given by
Eq.(\ref{difad}).

Solving this equation and using the $\delta$-function
approximation for ${\bf \Delta}(t)$ we obtain
${\cal
D}(t,\eta)=i\tau_{\rm 
e}^{-1}\,\hat{f}(\eta)\Lambda_{\eta}(t)$ with:
\widetext
\Lrule
\begin{eqnarray}
\label{Lambdalong}
\Lambda_\eta (t) = \int_{-\infty}^{t} d\xi\, C[{\bf
E}(\xi+\eta)-{\bf E}(\xi)]^{2}\,e^{-\gamma(t-\xi)}\,
\exp{\left\{
-\int_\xi ^{t} d\xi' \,C[{\bf E}(\xi'+\eta)-{\bf
E}(\xi')]^{2}
\right\}}, 
\end{eqnarray}
\Rrule
\narrowtext
\noindent
where $C[{\bf
E}(\xi+\eta)-{\bf E}(\xi)]^2$ is the short-hand notation for $[{\bf
E}(\xi+\eta)-{\bf E}(\xi)]_{i}C_{ij}[{\bf
E}(\xi+\eta)-{\bf E}(\xi)]_{j}$ and ${\bf
E}=-\partial {\bf A}/\partial t$ is the time-dependent electric field. We
also
re-installed the finite escape rate $\gamma$.
\subsection{Loose diffusons and the singularity of the quadratic response}
Eqs.(\ref{Lambda-fun}),(\ref{Lambdalong}) have similar structure:
$\Lambda_{\eta}(t)$ contains a quadratic in the external field pre-factor
multiplied by the exponential dephasing factor. Because of the quadratic 
in ${\bf A}$ pre-factor 
the loose
diffuson
does not arise in the linear response theory. However, if one considers
the quadratic in ${\bf A}$ response, the loose diffusons must be taken
into account while the
field-dependent dephasing should be neglected. In this approximation
we have:
\begin{equation}
\label{quad}
\Lambda_\eta (t) = \int_{-\infty}^{t} d\xi\, D[{\bf
A}(\xi+\eta)-{\bf A}(\xi)]^{2} \,e^{-\gamma(t-\xi)}
\end{equation}
and the similar expression in the quantum dot geometry.

Consider the harmonic pumping
 ${\bf A}={\bf
A}_{0}\cos(\omega
t)\,\theta(t)$
that is switched on at $t=0$. Then at a  time $t\gg \omega^{-1}$ the loose
diffuson averaged over the period is given by:
\begin{equation}
\label{lin}
\Lambda_{\eta}(t)=2 D {\bf A}_{0}^2 \,\sin^{2}\left(
\frac{\omega\eta}{2}\right)\,\frac{1-e^{-\gamma t}}{\gamma}.
\end{equation} 
One can see that in the limit $\gamma\rightarrow 0$ the loose diffuson is
{\it linear} in $t$.

We have already mentioned that the disorder average of the product of $k$ 
observables contains at most $k$ loose diffusons. Were the linear in $t$ 
growth
in Eq.(\ref{lin}) unrestricted, this would mean that the typical value
of a mesoscopic observable grows linearly with the running time. For a
particular case of the direct current arising in a mesoscopic ring under
ac pumping this statement can be found in Ref.\cite{Kop}.

Another similar statement concerns the {\it steady-state regime} when the
limit
$t\rightarrow\infty$ is taken in Eq.(\ref{lin}) {\it prior} to the limit
$\gamma\rightarrow 0$. One could argue from Eq.(\ref{lin}) that the
typical value of a mesoscopic
observable in the steady-state is diverging as
$\gamma\rightarrow 0$! 

However, Eqs.(\ref{Lambda-fun}),(\ref{Lambdalong}) clearly show that both
statements are 
artefacts of the quadratic approximation. 
In fact because of the {\it field-induced dephasing} the quantity 
$\Lambda_{\eta}(t)$ defined by Eqs.(\ref{Lambda-fun}),(\ref{Lambdalong})
is always smaller than 1. 
What is really singular in the limit $\gamma\rightarrow 0$ is the {\it
quadratic response susceptibility}. However, it does not mean a diverging
mesoscopic quantity, since at $\gamma\rightarrow 0$ the region of validity
of the quadratic
in ${\bf A}$ approximation shrinks to zero.
\subsection{Loose diffusons and the electron energy distribution}
Note that originating from the retarded-advanced junction Eq.(\ref{RAj})
the loose
diffuson is proportional to the combination
$\hat{f}(\eta)\Lambda_{\eta}(t)$, where $\hat{f}(\eta)$ is essentially the
Fourier-transform of the Fermi distribution function. From the procedure
of building the loose diffuson it is clear that any diagram with the loose
diffuson has a parent-diagram without the loose diffuson. For a particular
case of the variance of persistent  current in a mesoscopic ring
the
diagrams of Fig.7 (or Fig.8) with one or two loose diffusons stem from the
diagram of
Fig.6a (or Fig.6b) that contains no loose diffusons.  One can check (see
Sec.V) that the sum of all diagrams of the given family is equivalent to
replacing $\hat{f}(\eta)$ in the parent diagram by:
\begin{equation}
\label{repl}
\hat{F}(t+\eta,t)=[1-\Lambda_{\eta}(t)]\,\hat{f}(\eta).
\end{equation}
Consider now the simplest case of the steady-state. It corresponds to the
ac pumping switched on at $t=-\infty$. Let us define the time average:
\begin{equation}
\label{Gamma}
\Gamma(\eta)=\left\{\matrix{D\overline{[{\bf A}(t+\eta)-{\bf A}(t)]^2} &
{\rm ring\;or\;dot\;with\;}\omega\gg E_{c}\cr
 & \cr
C\overline{[{\bf E}(t+\eta)-{\bf E}(t)]^2}& {\rm dot\;
with}\;\omega\ll E_{c}\cr}\right.
\end{equation}
Substituting the time average $\Gamma(\eta)$ for $D[{\bf A}(t+\eta)-{\bf
A}(t)]^2$ or $C[{\bf E}(t+\eta)-{\bf E}(t)]^2$ in
Eqs.(\ref{Lambda-fun}),(\ref{Lambdalong}) we obtain the function
$\hat{F}(t+\eta,t)=\hat{F}(\eta)$ that depends only on the difference of
its
arguments:
\begin{equation}
\label{Fren}
\hat{F}(\eta)=\frac{\gamma\,\hat{f}(\eta)}{\gamma+\Gamma(\eta)}.
\end{equation}
We conclude that the loose diffusons amount to the {\it renormalization
of the energy distribution function} in a parent diagram.

This observation is an example of a generic rule that the `loose
propagators' in a filed theory
can be {\it eliminated} by the proper choice of the initial state
(`vacuum')
that in kinetics includes also the energy distribution function.
Eq.(\ref{Fren}) gives the form of this distribution for an open dynamical
system with no intrinsic relaxation and with the electron escape rate
$\gamma$. For the particular case of a harmonic pumping in the quantum 
dot geometry Eq.(\ref{Fren}) has been established in Ref.\cite{VAA}.

The renormalized energy distribution function Eq.(\ref{Fren}) retains the
property $F(E=\pm\infty)=\pm 1$ of the equilibrium distribution
$f(E)=\tanh(E/2T)$. This follows from the fact obvious from
Eq.(\ref{Gamma}) that $\Gamma(\eta\rightarrow 0)=0$. However, the form
of the energy distribution is different from $\tanh(E/2T)$
and strongly depends on the spectral content of the pumping field. For low
bath
temperature $T$ it contains at least two energy scales: the bath
temperature
$T$ and an additional scale $T_{*}$ set by the condition
$\Gamma(1/T_{*})=\gamma$. For a harmonic pumping  
$\Gamma(\eta)=4\gamma {\cal N}\,\sin^2 (\omega\eta/2)$, 
where:
\begin{equation}
\label{N}
{\cal N}=\left\{\matrix{\frac{D \overline{{\bf E}(t)^2}}{\gamma\omega^2} &
{\rm
ring\;or\;dot\;with\;}\omega\gg E_{c}\cr\
 & \cr
\frac{C \overline{{\bf E}(t)^2}}{\gamma} & {\rm dot\;with}\; \omega\ll
E_{c}} \right.
\end{equation}
For a strong pumping with ${\cal N}\gg 1$ one finds
\begin{equation}
\label{T*}
T_{*}=\omega \sqrt{{\cal N}}.
\end{equation}
This result \cite{VAA} corresponds to the diffusion in the energy space
with
$T_{*}$
being the displacement and
${\cal N}$ being the number of random-walk steps for the time
$\gamma^{-1}$ 
each with emitting or absorbing the energy $\omega$. In this case the
inverse Fourier transform $F(E)$ of $\hat{F}(\eta)$ is dominated by the
time intervals near zeros of $\Gamma(\eta)$ and we obtain:
\begin{equation}
\label{FE}
F(E)=\frac{1}{2\sqrt{{\cal N}}}\sum_{k=-\infty}^{+\infty}\tanh\left(\frac{E-\omega 
k}{2T}\right) \,e^{-\frac{|k|}{\sqrt{{\cal N}}}}.
\end{equation}
At $T\ll \omega \ll T_{*}$ the function $F(E)$ changes from -1 to +1 over 
the scale $T_{*}$ which plays a role of the effective electron
temperature. However it has a fine structure of sharp small steps with the
width of the  
transition regions being equal to the bath temperature $T$ and the width
of the
plateaus being equal to $\omega$ (see Fig.5).
\begin{figure}[-b]
\centerline{
\epsfig{figure=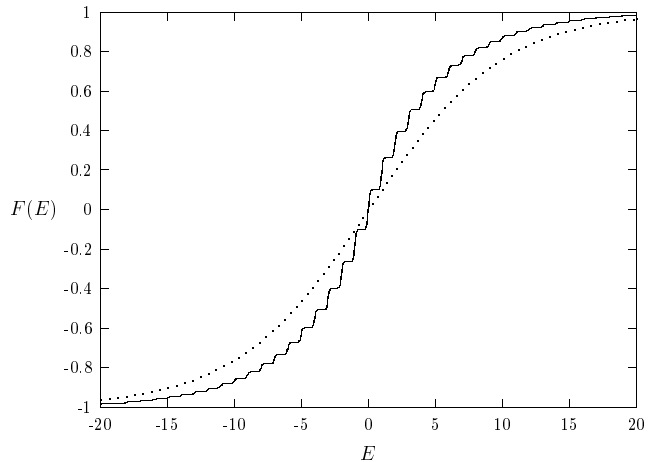,width=20pc,angle=0}
\vspace{5mm}
}
\caption{The anti-symmetric part $F(E)=1-2n(E)$ of the electron
energy
distribution $n(E)$ for
an open mesoscopic system of free electrons under harmonic pumping. Solid 
line corresponds to ${\cal N}=25$, $\omega=1$, $T=0.1$; the dotted line is
$\tanh(E/2T_{*})$.}
\end{figure}
In case of the white-noise pumping
$\Gamma(\eta)=2{\cal N}\gamma=const$ for all $\eta\neq 0$ and we formally
obtain
$T_{*}\rightarrow\infty$. The same result follows from
Eqs.(\ref{T*}),(\ref{N}) in the limit of a closed system
$\gamma\rightarrow 0$. 
On the diagrammatic level this manifests itself in the identity 
$\Lambda_{\eta}=1$ that holds in the limit $\gamma\rightarrow 0$ for all
$\eta\neq 0$ and causes a cancellation of all the diagrams in
the given family \cite{average}.

In the absence of dissipation or an electron escape the result that the
effective electron
temperature in the steady state $T_{*}=\infty$ is trivially correct but
is certainly
unphysical. It clearly
sets the limit of the dynamical approach and shows a necessity to take
account of an intrinsic dissipation in closed mesoscopic systems.
\subsection{Loose diffusons and the kinetic equation}
In this subsection we demonstrate that the function $\hat{F}(t_{1},t_{2})$
defined by Eq.(\ref{repl}) is
indeed the solution to the kinetic equation. For simplicity we consider
the case of the ring geometry and set $\gamma=0$.

To derive the kinetic equation we start with the left
and the
right
Dyson equations for the $2\times 2$ Greens function ${\underline G}$:
\begin{mathletters}
\label{DysonK}
\begin{eqnarray}
\label{DysonK-1} 
({\underline G}_0^{-1} - {\underline \Sigma})\otimes {\underline G} = 
\delta({\bf x}_1-{\bf x}_1^\prime), \\
\label{DysonK-2}
{\underline G} \otimes ({\underline G}_0^{-1} - {\underline \Sigma}) =
\delta
({\bf x}_1-{\bf x}_1^\prime).
\end{eqnarray}
\end{mathletters}
Here, the standard notations \cite{RS-1986} were adopted with
${\bf x}_k = ({\bf r}_k,t_k)$, and
\begin{eqnarray}
\label{inv-green}
G_0^{-1}({\bf x}) = i\frac{\partial}{\partial t} -
\xi_{{\bf p}= -i{\bf \nabla}_{\bf r} - {\bf A}(t)}.
\end{eqnarray}
 
As the external AC field is
a classical field, the components $\Sigma^{\rm K(R,A)}$ of the self energy
\begin{eqnarray}
\label{se}
{\underline \Sigma}({\bf r}t,{\bf r^\prime}t^\prime) =
\left(
\begin{array}{cc}
\Sigma^{\rm R} & \Sigma^{\rm K} \\
0 & \Sigma^{\rm A}
\end{array}
\right)
\end{eqnarray}
are
\begin{eqnarray}
\label{s-energy} 
\Sigma^{\rm K(R,A)} = ({\bf {\hat j} A})({\bf x}_1) G^{\rm K(R,A)}({\bf
x}_1,
{\bf x}_2) ({\bf {\hat j}  A})({\bf x}_2).
\end{eqnarray}
Subtracting the two Dyson equations Eq. (\ref{DysonK}) from one another,
and taking 
the Keldysh component of the result, one obtains:
\widetext
\Lrule
\begin{eqnarray}
\label{lhs=rhs}
\left[ G_0^{-1}({\bf x_1})-{G_0^{-1}}^*({\bf x_2})\right] G^{\rm K}({\bf
x_1},{\bf x_2})
=
\left[\Sigma^{\rm K}\otimes G^{\rm A} - G^{\rm R}\otimes \Sigma^{\rm K} +
\Sigma^{\rm R}\otimes G^{\rm K} - G^{\rm K}\otimes \Sigma^{\rm A}\right]
({\bf x_1},{\bf x_2}).
\end{eqnarray}
\Rrule
\narrowtext
\noindent
Before doing the next step we introduce the {\it time-dependent} energy
distribution function $n(E,t)=(1/2)[1-f(E,t)]$, where $f(E,t)$ is related
to the Keldysh component of the matrix Green function:
\label{Keldysh-ansatz}
\begin{eqnarray}
\label{ka-1}
{\bf G}^{\rm K}({\bf r}_1 t_1, {\bf r}_2 t_2) &=& \int d\tau
\left[
F(t_1,\tau) {\bf G}^{\rm R}({\bf r}_1 \tau, {\bf r}_2 t_2) \right.
\nonumber \\
&-& \left. {\bf G}^{\rm A}({\bf r}_1 t_1, {\bf r}_2 \tau) F(\tau,t_2)
\right], \\
\label{ka-2}
f(E,t) &=& \int \frac{d\eta}{2\pi} e^{-iE\eta}
F\left(t+\frac{\eta}{2},t-\frac{\eta}{2} \right).
\end{eqnarray}

Now we substitute Eqs. (\ref{s-energy}) and (\ref{ka-1})
into
Eq. (\ref{lhs=rhs}), and perform the disorder averaging.
Using the identity:
\begin{eqnarray}
\label{df-fk}
F(t_1,t_2) = \frac{i}{2\pi\nu} \sum_{\bf p} \langle G^{\rm K}
({\bf p};t_1,t_2)\rangle,
\end{eqnarray}
we arrive at the equation
\begin{equation}
\label{f-eq}
\left(
\frac{\partial}{\partial t_1} + \frac{\partial}{\partial t_2}
\right) F(t_1,t_2) = -D \left[
{\bf A}(t_1) - {\bf A}(t_2)
\right]^2 F(t_1,t_2).
\end{equation}
The latter is easy to solve by introducing the function
${\tilde F}(t,\eta)= F(t+\eta/2,t-\eta/2)$, with
$t=(t_1+t_2)/2$
and $\eta = t_1 - t_2$ being the Wigner variables:
\begin{equation}
\label{df-wigner}
\frac{\partial}{\partial t} {\tilde F}(t,\eta) = -D \left[
{\bf A}\left(t+ \frac{\eta}{2}\right) - {\bf A}\left(t -
\frac{\eta}{2}\right)
\right]^2 {\tilde F}(t,\eta).
\end{equation}
In accordance with Eq. (\ref{ka-2}), the variable $t$ has a clear
meaning of the global running time. We supplement Eq.(\ref{df-wigner})
by the initial condition ${\tilde F}(t=-\infty,\eta)={\hat f}(\eta)$ to
end
up with:
\begin{equation}
\label{answer-4-df}
F(t+\eta,t) = {\hat f}(\eta) \exp\left\{
-D\int_{-\infty}^{t} d \xi \left[
{\bf A}(\xi+\eta) - {\bf A}(\xi)
\right]^2
\right\}.
\end{equation}
Comparison of Eq. (\ref{answer-4-df}) with 
(\ref{Lambda-fun}) (at $\gamma=0$)
shows that the function $F(t_{1},t_{2})$ is identical to the function
$\hat{F}(t_{1},t_{2})$ defined by Eq.(\ref{repl}) which follows from the
evaluation of diagrams with loose diffusons. 

This proves that the loose diffuson $\Lambda_{\eta}(t)$ determines
the {\it time-dependent} electron energy distribution function
$n(E,t)=(1/2)\,[1-f(E,t)]$:
\begin{equation}
\label{tdep}
f(E,t)=\int \frac{d\eta}{2\pi}\,e^{-i E
\eta}\,\hat{f}(\eta)\,[1-\Lambda_{\eta}(t)],
\end{equation}
where $\Lambda_{\eta}(t)$ is given by
Eqs.(\ref{Lambda-fun}),(\ref{Lambdalong}).
\section{Fluctuations of persistent current in mesoscopic metallic rings
in and out of equilibrium}
For illustration purposes we consider in this section how the general 
formalism
described above works in the particular problem of the persistent current
\cite{BIL-1983} in mesoscopic rings pierced by a constant magnetic
flux $\phi$ and subject to ac pumping. Since the
disorder-averaged  persistent current of non-interacting electrons
considered in the grand-canonical ensemble is 
exponentially small \cite{CRG-1989} we concentrate on the {\it 
mesoscopic fluctuations} of persistent current at temperatures $T\gg
1/\tau_{\phi}$ where $1/\tau_{\phi}$ is the total dephasing rate including
that of the ac pumping. This condition allows to neglect dephasing
everywhere but in the loose diffusons which describe the evolution of
the electron energy distribution under the ac pumping. We 
will show by straightforward diagrammatic calculations 
that Eq.(\ref{repl}) indeed holds if all diagrams of the given family are
taken into account. This illustrates how the diagrammatic technique takes
care of the correct electron energy distribution in the non-eqilibrium
problem. 
\subsection{Equilibrium fluctuations of persistent currents}
We start with the equilibrium fluctuations of persistent currents in
order to
specify the {\it parent diagrams} for the problem considered. 

Following a standard
route, we express the persistent current  in terms of
exact retarded and
advanced electron Green's functions $G_{0}^{R(A)}$:
\begin{eqnarray}
\label{pc-gf}
I_{\rm PC} = i \int \frac{dE}{2\pi} f(E) {\rm Tr}
\left[
{\hat j}_\alpha
(
G_{0}^{R}(E) - G_{0}^{A}(E)
)
\right].
\end{eqnarray}
Then the variance of the persistent current fluctuations is given by:
\begin{eqnarray}
\label{fluct-greenfunction}
\langle I_{\rm PC}^2\rangle &=& 2\int
\frac{dE}{2\pi}\int\frac{dE^\prime}{2\pi}
f(E)f(E^\prime) \nonumber \\
&\times& \left\langle {\rm Tr}\left\{ {\hat {\bf j}} G_{0}^{R}(E)\right\}
{\rm Tr}\left\{ {\hat {\bf j}} G_{0}^{A}(E^\prime)\right\}
\right\rangle. 
\end{eqnarray}
To perform the impurity averaging in Eq. (\ref{fluct-greenfunction}),
it is convenient
to use a representation \cite{GLK-1979,H-1981} in which slow
diffusion and fast ballistic modes are explicitly separated from each
other.   
The lowest
order
(single--loop) diagrams contributing to the persistent current
fluctuations 
are shown   
in Fig. 6.
There, wavy lines correspond to Diffuson (${\cal D}_k$) or
Cooperon
(${\cal C}_k$) propagators, while
triangles and a square (whose edges correspond to the average Green
functions)
represent the electron motion on the ballistic scale. On the diffusion
scale, the latter are reduced to certain constants. 
\begin{figure}[-b]
\centerline{
\epsfig{figure=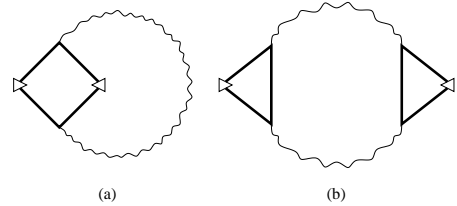,width=14pc,angle=0}
\vspace{5mm}}
\caption{
Local (a) and nonlocal (b)  single--loop diagrams for
the
equilibrium
persistent current fluctuations. Triangular vertices mark the current
operators
${\hat j}_\alpha$. These are the {\it parent diagrams} for the persistent
current fluctuations under ac pumping.
}
\label{fig1}
\end{figure}
Applying conventional
rules
(e.g., see Ref. \cite{AKL-1991}), we convert the diagrams in Figs. 6(a)
and 6(b) to
\widetext
\Lrule
\begin{mathletters}
\label{pc-dc}
\begin{eqnarray}
\label{pc-single-dc}
\langle {I_{\rm PC}^2}\rangle _{\rm l} &=&
4\pi \nu D \tau_{\rm e}^2 e^2 \int d\eta  {\hat f}^2(\eta)
\sum_{m}
[
{\cal C}_{k_{m-2\phi/\phi_0}}(\eta)-{\cal D}_{k_m}(\eta)
], \\
\label{double-cd}   
\langle {I_{\rm PC}^2} \rangle _{\rm nl} &=
& (4\pi \nu D \tau_{\rm e}^2)^2 e^2 \int d\eta
{\hat f}^2(\eta)
\int d\xi
\sum_{m}
[
k_m^2{\cal D}_{k_m}(\xi){\cal D}_{k_m}(\eta -\xi)
-
k_{m-2\phi/\phi_0}^2{\cal C}_{k_{m-2\phi/\phi_0}}
(\xi){\cal C}_{k_{m-2\phi/\phi_0}}
(\eta -\xi)
].
\end{eqnarray}
\end{mathletters}
\Rrule
\narrowtext
\noindent  
Here
$k_m=(2\pi/L)m$ with $m$
running over all integers represents the spectrum of diffusion modes
allowed for
conductor with the ring topology, $L$ being the circumference.
In Eqs. (\ref{pc-dc}),
the contributions of single-- and double--diffuson (cooperon) diagrams
have
been
singled out in the time domain, in which
\begin{eqnarray}
\label{diff-coop}
{\cal D}_k(t) \equiv {\cal C}_k (t) = \frac{\theta(t)}{2\pi\nu \tau_{\rm
e}^2}
\exp(-Dk^2 t).
\end{eqnarray}
 
In fact, Eq. (\ref{diff-coop}) allows us to effectively compactify the
nonlocal
diagrams, Fig. 1(b), so that the total fluctuations $\langle I_{\rm PC}^2
\rangle$
are solely
expressed in terms of the contributions of the local diagrams, Fig. 1(a):
\begin{eqnarray}
\label{compact}
\langle I_{\rm PC}^2 \rangle &=& 8\pi\nu D \tau_{\rm e}^2 e^2
\int d\eta {\hat f}^2(\eta)
\left(
D\frac{\partial}{\partial D} + \frac{\partial}{\partial D} D
\right)   
\nonumber \\
&\times& \sum_{k_m}
[{\cal C}_{k_m-2\phi/\phi_0}(\eta) - {\cal D}_{k_m}(\eta)].
\end{eqnarray}
The fluctuations of persistent currents, Eq. (\ref{compact}) are
manifestly
periodic
in the flux $\phi$, with the period $\phi_0/2$. This can explicitly be
displayed by performing the re-summation in Eq. (\ref{compact}) using
the
Poisson formula:
\begin{eqnarray}
\label{pc-fluctuations}
\langle I_{\rm PC}^2 \rangle &=& \sum_{n=1}^
\infty \langle I_n^2 \rangle \sin ^2\left(
2\pi n \frac{\phi}{\phi_0}
\right)
\end{eqnarray}  
with
\begin{eqnarray}
\label{I-harmonics}
\langle I_n^2 \rangle &=& \frac{4C^{2} n^2}{\pi^{1/2}}
\left( \frac{e}{\tau_{\rm D}} \right)^2 \,{\tilde T}^{2}
\int_0^\infty \frac{dx}{x^{3/2}}
\frac{e^{-n^2/(4x)}}{\sinh^2(\pi {\tilde T} x)},
\end{eqnarray}  
where ${\tilde T}= T/E_c$ denotes the electron temperature measured
in the units of the Thouless energy $E_c = 1/\tau_{\rm D}$. For
generality we also introduce the
coefficient $C$ which is equal to 1 for the case of potential disorder
(orthogonal ensemble) considered here and $C=1/2$ for the strong
spin-orbit interaction
(symplectic ensemble). 

Eq.(\ref{I-harmonics}) is in a complete correspondence with the earlier
results
obtained in Refs.\cite{vOR,Mont}.

\subsection{Effect of ac pumping on the persistent current fluctuations}
Now let us assume that a time-dependent circular field ${\bf A}(t)$ [see
Fig.3a] is
applied to the mesoscopic ring and consider the persistent current
fluctuations under ac pumping. The dc current in a ring (overline means
the time averaging)
$I_{dc}=\overline{I^{(1)}(t)+I^{(2)}(t)}$
can be
found 
using Eqs.(\ref{I-expansion})-(\ref{ra}) [see also Fig.2]:
\begin{equation}
\label{I1}
I^{(1)}(t)=i\int d\eta \,\hat{f}(\eta)\,{\rm Tr}\left\{\hat{{\bf
j}}{\bf
G}^{R}(t,t-\eta)-\hat{{\bf j}}{\bf
G}^{A}(t+\eta,t) \right\},
\end{equation}
\begin{eqnarray}
\label{I2}
I^{(2)}(t)&=&i\int dt_{1}\int d\eta\,\hat{f}(\eta)\,[{\bf
A}_{\alpha}(t_{1}+\eta)-{\bf A}_{\alpha}(t_{1})] \nonumber \\ &\times& 
{\rm
Tr}\left\{\hat{{\bf
j}}{\bf
G}^{R}(t,t_{1}+\eta)\hat{{\bf j}}_{\alpha}{\bf
G}^{A}(t_{1},t)\right\},
\end{eqnarray}
where ${\bf G}^{R,A}$ are exact electron Green's functions in the persence
of ac pumping Eq.(\ref{ra}).

One can check that for equilibrium electron Green's functions  
Eq.(\ref{I1}) reduces to Eq.(\ref{pc-gf}). The contribution Eq.(\ref{I2})  
is present only under ac pumping. It describes two principally different
effects of ac pumping. One is the {\it rectification} of ac field
discussed in Refs.\cite{DC,krav3}. This effect is similar to the {\it
photovoltaic effect} in a single-connected geometry \cite{FKh,VAA}.
Another one is the heating by ac field which 
we will study starting from the simplest {\it zero order} in the pump
field parent diagrams of Fig.6. The heating effect in rectification (or
photovoltaic effect) can be studied in a similar way \cite{VAA} starting
from the parent diagrams of the {\it second order} in the pump field
\cite{FKh,VAA,YK}.

The daughter-diagrams with loose diffusons which arise after disorder
averaging and correspond
to the parent-diagram of Fig.6a. are given in Fig.7.
We stress that although the parent diagram of Fig.6a arises as the result
of the disorder averaging of $\langle I_{1}^2 \rangle$, the
daughter-diagrams always involve $I_{2}$ that contains the
retarded-advanced junction and allows to build the loose diffuson.
In a similar way we get the non-local daughter-diagrams that correspond
to the parent diagram of Fig.6b.
\widetext
\begin{figure}
\centerline{
\epsfig{figure=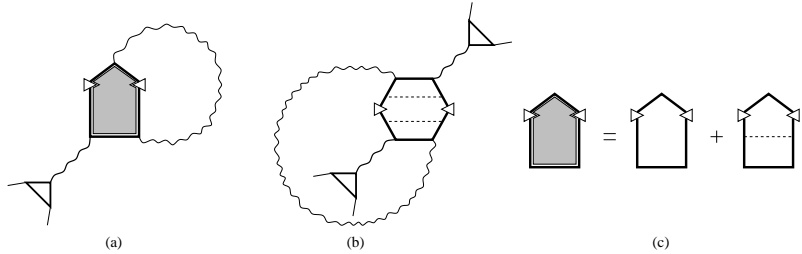,width=25pc,angle=0}
\vspace{5mm}
}
\caption{
Local single--loop daughter-diagrams with one (a) or two (b)
loose diffusons contributing to the
non-equilibrium
fluctuations of persistent current. The diagram (a) arises from the  
disorder average $\langle I_{1}(t)I_{2}(t') \rangle$ and the diagram (b)
arises
 from the
disorder average $\langle I_{2}(t) I_{2}(t') \rangle$.
They are complementary to the parent-diagram of Fig. 6(a).
Hikami box of the diagram (a) is detailed in (c).
}
\label{fig4}
\end{figure}
\narrowtext
\noindent
\widetext
\begin{figure}[-b]
\centerline{
\epsfig{figure=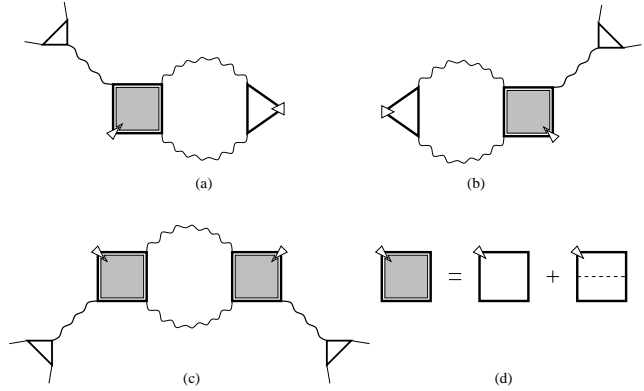,width=20pc,angle=0}
\vspace{5mm}
}
\caption{   
Nonlocal single--loop daughter-diagrams: the diagrams (a) and (b) arise
from the disorder average of $\langle I_{1}(t)I_{2}(t')\rangle$ while the
diagram (c)
arises from the disorder average $\langle I_{2}(t)I_{2}(t')\rangle$.
These are
complementary to
the parent-diagram of Fig. 6(b). The corresponding Hikami box (d) is also
shown.
}
\label{fig5}
\end{figure}
\narrowtext
\noindent

Calculating the diagrams of Fig.7 and Fig.8 and assuming $T\gg\gamma,
1/\tau_{\phi}$ we neglect both dephasing and
electron escape in the loop diffusons/cooperons and adopt
Eq.(\ref{diff-coop}) to describe them. Then summing up all the
diagrams of Fig.6-8 we arrive at the expression for the disorder average
$\langle I_{PC}(t) I_{PC}(t')\rangle$ which is exactly the same as 
Eq.(\ref{compact}) with $\hat{f}^{2}(\eta)$ replaced by
$\hat{f}^{2}(\eta)\,(1-\Lambda_{\eta}(t))(1-\Lambda_{\eta}(t'))$. This
leads to the ansatz Eq.(\ref{repl}).

In particular, for a harmonic pumping with the frequency
$\omega=\tilde{\omega} E_{c}$ we obtain from
Eq.(\ref{Fren}):
\begin{eqnarray}
\label{harmfluc}
\langle I_n^2 \rangle &=& \frac{4C^{2} n^2}{\pi^{1/2}}
\left( \frac{e}{\tau_{\rm D}} \right)^2 \,
\int_0^\infty \frac{dx}{x^{3/2}}
e^{-n^2/(4x)}\, F^{2}(x),
\end{eqnarray}
where
\begin{equation}
\label{Gamharm}
F(x)=
\frac{\tilde{T}}{\sinh(\pi\tilde{T}x)}\,\frac{1}{\left[1+4{\cal N}\,\sin^2
\left(\frac{\tilde{\omega}x}{2}\right)\right]}.
\end{equation}
Here we assume that the ring is connected to the electron reservoir by a 
{\it passive lead} [see Fig.10b] which results in a finite electron
escape rate
$\gamma$ and allows to reach a steady-state regime. The escape rate 
enters the constant ${\cal N}$ in Eq.(\ref{N}) that is equal to the number
of
absorption/emission events for  the escape time and thus 
describes the
pumping strength. 

At $T_{*}=\omega\sqrt{{\cal N}}\gg E_{c}$ the
variance of
persistent current fluctuations is stongly suppressed 
but still significantly depends on the bath temperature $T$
even if $T \ll T_{*}$ [see Fig.9]. 
\begin{figure}[-b]
\centerline{
\epsfig{figure=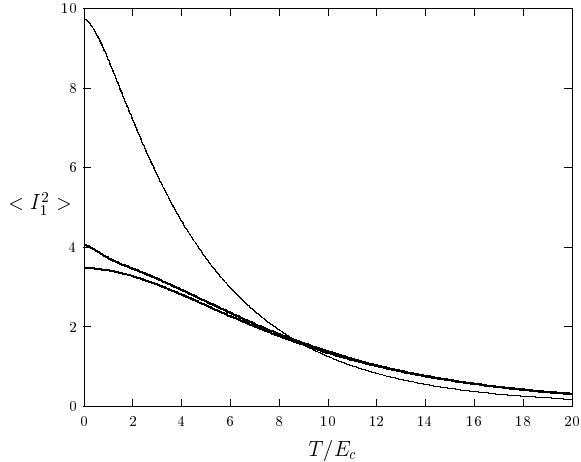,width=18pc,angle=0}
\vspace{5mm}
}
\caption{Temperature dependence of the variance of the first harmonic of
persistent current in units of $(e/\tau_{D})^2$: the equilibrium
temperature dependence (upper curve) and two close curves (blown up by a
factor of 5) for
the non-equilibrium case with $T_{*}=20 E_{c}$ corresponding to 
$\tilde{\omega}=10$, ${\cal N}=4$ and
$\tilde{\omega}=2$, ${\cal N}=100$. The temperature dependence is
significant in
the
non-equilibrium case even for $T\ll T_{*}$.}
\end{figure} 
For instance at
$\omega\ll E_{c}$ and $T_{*}=\omega\sqrt{{\cal N}}\gg E_{c}$ we obtain 
\begin{equation}
\label{EXAMPLE}
\langle I_n^2 \rangle \propto 
\left(\frac{E_{c}}{T_{*}}\right)^4 \,
\int_0^\infty \frac{dx}{x^{11/2}}
e^{-n^2/(4x)}\, \frac{\tilde{T}^2}{\sinh^{2}(\pi \tilde{T}x)}
\end{equation}
Eq.(\ref{EXAMPLE}) shows the dependence on the bath temperature
$T=\tilde{T}E_{c}$ which is of the same type as in the
absence of
pumping Eq.(\ref{I-harmonics}), only the magnitude of fluctuations 
decreases by the factor of $(E_{c}/T_{*})^{2} \ll 1$.
So the overall width $T_{*}$ and the small steps in the electron energy
distribution of Fig.5 manifest themselves in the variance of
persistent current fluctuations. 

\section{ dc conductance under ac pumping}
In this section we consider the dc conductance in a
mesoscopic system under ac pumping. This problem has been recently addressed in the
work by Pedersen and B\"uttiker \cite{butt}.
Here we neglect the electron interaction and focus on the
effect of heating by the ac field. It is well known that the mesoscopic
conductance fluctuations are temperature dependent and decrease
when the size of the system $L \gg L_{T}$ where
$L_{T}=\sqrt{D/T}$. The question we address here is whether or not the
effective temperature $T_{*}$ is what should stand for
the bath temperature $T$ in the expression for $L_{T}$ under ac pumping.
The answer is not obvious in the geometry of an open quantum dot
connected by the leads to electron reservoirs [ see Fig.10a,b].
The point is that there are {\it several} different
electron energy distributions in such a problem. The electron
energy distribution in each reservoir is supposed to be the equilibrium
Fermi
distribution with a certain chemical potential and temperature. 
In addition, there is the non-equilibrium electron energy distribution
inside the
dot under ac pumping. 

We define the {\it Landauer conductance} as the
linear dc
 current response to the {\it difference of the chemical potentials}
between
two different reservoirs with the same temperature [Fig.10a].

Another experimental situation corresponds to measurements of the linear
dc current response
to the {\it perturbation of the system's Hamiltonian} caused by the
constant electric field inside the mesoscopic system. The corresponding 
response function will be referred to as the {\it
Kubo conductance}.
It can be realized as a current response in a ring that is
pierced by a magnetic flux \cite{butt2} [see Fig.10b].  
\begin{figure}[-b]
\centerline{
\epsfig{figure=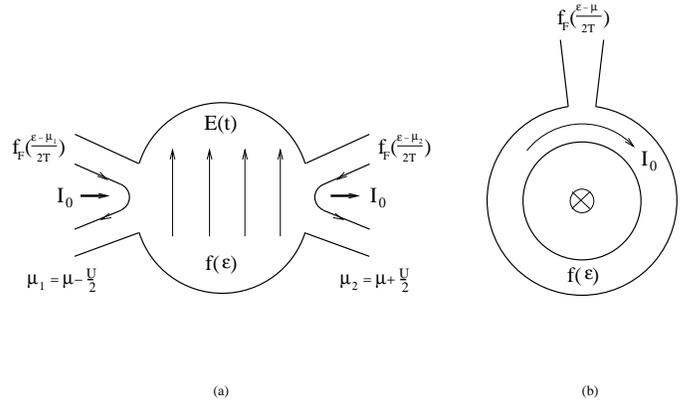,width=21pc,angle=0}
\vspace{5mm}  
}
\caption{Experimental geometries for the Landauer (a) and Kubo (b)  
conductances. In the case of  Landauer conductance the dc current $I_{0}$
is the response to the chemical potential difference $\mu_{1}-\mu_{2}$
between two reservoirs which
enter through the Fermi-distributions of the {\it incoming}
electrons. In the case of Kubo conductance the dc current
$I_{0}$ is the
response to the perturbation $\hat{j}E_{0}t$ of the system's Hamiltonian.
The {\it passive lead} connects the mesoscopic ring to the electron
reservoir with the equilibrium distribution of electrons.}
\end{figure}
The flux is supposed to contain
two parts: one
is growing linearly
with time and causes the dc electric field. Another one produces the
high-frequency pumping. 

We will show that these two cases are drastically different (the 
same conclusion has been reached by Vavilov and Aleiner \cite{AleinPr}). The
Landauer
conductance
of non-interacting electrons is {\it insensitive} to heating and depends only
on the bath temperature $T$. At the same time
the non-equilibrium electron energy
distribution in the mesoscopic ring does matter for the {\it Kubo
conductance}. In particular the
mesoscopic fluctuations of the Kubo conductance should feel the effective
temperature $T_{*}$
rather than the temperature of the electron reservoir $T$ which the
mesoscopic
ring is connected to by the {\it passive lead}.
\subsection{Landauer conductance in terms of  electron Green's
functions in the time domain}

In order to prove the statement on the
absence of
sensitivity of the Landauer conductance to heating produced by ac pumping
in the dot and to see the key difference between the Landauer and the Kubo
conductances we re-derive the Landauer conductance to allow for an
arbitrary ac pumping. 
To derive the expression for the Landauer conductance in terms of the
electron Green's functions we proceed along the route 
used in Ref.\cite{VAA}. 
\subsubsection{Formulation of the Landauer conductance}

For simplicity we consider a
zero-dimensional dot described by the $N\times N$ random {\it
time-dependent} matrix $H_{ij}(t)$
connected to two
perfect semi-infinite ($x<0$) leads each containing $M$ channels labeled
by $\alpha$.
 There is neither 
disorder nor electron  interaction with each other or with an external ac
field 
inside the leads. Therefore  the matrix Green's function $\underline
{G}_{++,\alpha\beta}(t-t',x-x')$ of the Keldysh
technique Eq.(\ref{Green-Keldysh}) {\it for 
incoming electrons} in the leads  depends only
on the
difference of time
and coordinates \cite{outgoing}. Moreover, the {\it incoming electrons}
are supposed to be
in equilibrium at the bath temperature $T$ and the chemical potential
$\mu+\delta\mu$,
where $\delta\mu=\pm  U/2$ differs in sign for 
the leads 1 and 2.   
Then the Keldysh component $G_{++,\alpha\beta}^{K}(E,x-x')$ of
the incoming electron Green's function $\underline{      
G}_{++,\alpha\beta}(E,x-x')$ in the energy-coordinate representation
takes the form [cf. Eq.(\ref{GK})]: 
\begin{eqnarray}
\label{Kk}
&
&G^{K}_{++,\alpha\beta}(E,x-x')=\tanh\left(\frac{E-\delta\mu_{\alpha}}{2T}
\right)\\ \nonumber &\times&[G_{++,\alpha\beta}^{R}(E,x-x')-
G_{++,\alpha\beta}^{A}(E,x-x')].
\end{eqnarray}
The retarded and advanced components for incoming electrons
$G^{R(A)}_{++,\alpha\beta}(t-t',x-x')$ in the leads are given
by:
\begin{eqnarray}
G^{R}_{++,\alpha\beta}&=&
i\theta(t-t')\,\delta_{\alpha\beta}\,
\delta(v_{F}(t-t')-(x-x'))\\ 
\label{RA-1}
G^{A}_{++,\alpha\beta}&=&
-i\theta(t'-t)\,\delta_{\alpha\beta}
\,\delta(v_{F}(t-t')-(x-x')).
\label{RA-2}
\end{eqnarray}
Here we linearized the Schr\"odinger equation near the Fermi momentum and 
introduced right (incoming) and left (outgoing) movers
$\Psi_{\pm}^{(\alpha)}(x,t)=A_{\pm}\,e^{\pm
ikx-iv_{F}kt}\,e^{\pm ip_{F}x}$. Then the wave function in the leads
$(x<0)$ is:
\begin{equation}
\label{wf}
\Psi^{(\alpha)}(x,t)=\Psi_{+}^{(\alpha)}(x,t)+\Psi_{-}^{(\alpha)}(x,t).
\end{equation}

The Schr\"odinger equation for electron states inside the dot
$\psi_{n}(t)$ coupled
to electron states in the leads $\Psi_{\pm}^{(\alpha)}(x,t)$ is taken in
the form:
\begin{equation}
\label{Sch}
\sum_{m}[i\delta_{nm}\partial_{t} -
H_{nm}(t)]\,\psi_{m}(t)=\sum_{\alpha}W^{\dagger}_{n\alpha}\,
\Psi^{(\alpha)}(0,t),
\end{equation} 
where $W_{\alpha n}$ is the $2M\times N$ coupling matrix.

Neglecting electron-electron interaction we introduce the {\it linear}
boundary conditions at $x=0$:
\begin{equation}
\label{bcon}
-iv_{F}\,[\,\Psi_{+}^{(\alpha)}(0,t)- 
\Psi_{-}^{(\alpha)}(0,t)]=\sum_{m} W_{\alpha 
m}\,\psi_{m}(t),
\end{equation}
For massive leads with the semiclassical
electron motion, electrons adiabatically turn back at $W_{\alpha
m}\rightarrow 0$ acquiring only a certain phase which may be included
into the definition of $\Psi_{\pm}^{(\alpha)}$. 

Equations Eqs.(\ref{Sch}),(\ref{bcon}) generate the corresponding
equations for the matrix Green's functions:
\begin{eqnarray}
\label{Sch1} 
& &\sum_{m}[i\delta_{nm}\partial_{t} -
H_{nm}(t)]\,\underline{G}_{\pm,m\beta}(t,t';x')\\ \nonumber &=&
\sum_{\alpha}W^{\dagger}_{n\alpha}\,
[\underline{G}_{+\pm,\alpha\beta}(t,t';-x')+\underline{G}_{-\pm,\alpha\beta}
(t,t';-x')],
\end{eqnarray}
\begin{eqnarray}
\label{bcon1}
& &\sum_{m}
W_{\alpha
m}\,\underline{G}_{\pm,m\beta}(t,t';x')=-iv_{F}\\ \nonumber
&\times&[\,\underline{G}_{+\pm,\alpha\beta}(t,t';-x')-
\underline{G}_{-\pm,\alpha\beta}(t,t';-x')],
\end{eqnarray}
where the label $+(-)$ corresponds to the incoming (outgoing) electrons in 
the leads and $\underline{G}_{\pm,m\beta}$ is the `cross' Green's function
of an incoming (outgoing) electron in a channel $\beta$ in the leads
and an electron at a site $m$ in the dot. 

According to
Eq.(\ref{I-definition}) the current $I_{1,2}(t)$ in the leads 1 or 2 is
given
by the
Keldysh components $G^{K}_{++,\alpha\alpha}(t,t;0)$ and
$G^{K}_{--,\alpha\alpha}(t,t;0)$ of the incoming and outgoing electrons:
\begin{equation}
\label{curr}
I_{1,2}=iv_{F}\sum_{\alpha\in 
1,2}[G^{K}_{++,\alpha\alpha}(t,t;0)-G^{K}_{--,\alpha\alpha}(t,t;0)].
\end{equation} 
\subsubsection{Analytical structure of the scattering matrix}
It is possible \cite{VAA} to
express the Keldysh component
$G^{K}_{--,\alpha\beta}(t,t';0)$ of the outgoing electrons in terms
of the known Keldysh component $G^{K}_{++,\alpha\beta}(t,t';0)$ for the
incoming electrons using the time-dependent scattering matrix
$S_{\alpha\beta}(t,t')$:
\begin{eqnarray}
\label{SM}
& & G^{K}_{--,\alpha\beta}(t,t';0)=\\ \nonumber
&=&\int dt_{1}\int
dt_{2}\,S_{\alpha\gamma}(t,t_{1})\,G^{K}_{++,\gamma\delta}(t_{1}-t_{2};0)
\,S^{\dagger}_{\delta\beta}(t_{2},t').
\end{eqnarray} 
It is crucial for us that the scattering matrix $S_{\alpha\beta}(t,t')$
involves only the {\it retarded} component ${\bf G}^{R}_{mn}(t,t')$ of the
electron Green's function {\it inside the dot} and the Keldysh component
${\bf G}^{K}_{nm}(t,t')$ drops out of the scattering matrix \cite{VAA}:
\begin{equation}
\label{SCM}
S_{\alpha\beta}(t,t')=\delta_{\alpha\beta}\,\delta(t-t')-2iv_{F}^{-1}\,
W_{\alpha n}\,{\bf G}^{R}_{nm}(t,t')\,W^{\dagger}_{m\beta}.
\end{equation}
We remind that only
the Keldysh component ${\bf G}^{K}_{nm}(t,t')$ contains an information
on the
electron energy distribution  inside the dot. It
describes the {\it real} transitions between energy levels in the dot
which are constrained by the energy conservation law and the Pauli
principle. In contrast to that
the retarded component ${\bf G}^{R}_{mn}(t,t')$ describes {\it virtual}
transitions where no energy conservation applies and thus the energy
distribution function is irrelevant.

Formally the fact that the scattering matrix is independent of ${\bf
G}^{K}_{mn}(t,t')$ follows from Eqs.(\ref{Sch1}),(\ref{bcon1}). Let us
express $\underline{G}_{-+,\alpha\beta}$ through
$\underline{G}_{++,\alpha\beta}$ and $\underline{G}_{+,m\beta}$ using
Eq.(\ref{bcon1}) and substitute it into Eq.(\ref{Sch1}). As the
result of the transformation 
$2 \underline{G}_{++,\alpha\beta}$ appears in the r.h.s. of  
Eq.(\ref{Sch1}) and
the matrix Hamiltonian for electrons in the
dot acquires an
imaginary part $H_{nm}\rightarrow H^{R}_{nm}$, where:
\begin{equation}
\label{ImH}
H^{R,A}_{nm}=H_{nm}\mp iv_{F}^{-1}\,(W^{\dagger}W)_{nm}.
\end{equation} 
According to the rules of the Keldysh technique the
generic solution $G^{K}_{+,m\beta}$ to the transformed Eq.(\ref{Sch1})
involves both
the combination ${\bf
G}^{R}_{mn}W^{\dagger}_{n\alpha}G^{K}_{++,\alpha\beta}$ and the
combination ${\bf 
G}^{K}_{mn}W^{\dagger}_{n\alpha}G^{A}_{++,\alpha\beta}$,
where $\underline{\bf G}$ is the marix Green's function for electrons in
the dot, for instance: 
\begin{equation}
\label{dotGF}
{\bf
G}^{R}_{nm}(t,t')=\left[i\partial_{t}-H^{R}(t) 
\right]_{nm}^{-1}.
\end{equation}
However, in this
particular scattering problem we have:
\begin{equation}
\label{A0}
G^{A}_{++,\alpha\beta}(t,t';x-x'>0)=0.
\end{equation}
Eq.(\ref{A0}) follows immediately from the $\delta$-function and the
$\theta$- function structure of Eq.(\ref{RA-2}). Therefore we obtain
from Eqs.(\ref{Sch1}),(\ref{bcon1}):
\begin{eqnarray}
\label{+-}
G^{K}_{-+,\alpha\beta}(t,t';0)=\int
dt_{1} S_{\alpha\gamma}(t,t_{1}) G^{K}_{++,\gamma\beta}(t_{1},t';0).
\end{eqnarray}
with $S_{\alpha\gamma}(t,t_{1})$ given by Eq.(\ref{SCM}) which contains
only the retarded component ${\bf G}^{R}_{nm}$ of the electron Green's
function in the dot.

In a similar way one can express $G^{K}_{--,\alpha\beta}(t,t';0)$
through $G^{K}_{-+,\alpha\beta}(t,t';0)$ using the Hermitean conjugated
equations Eq.(\ref{Sch1}),(\ref{bcon1}) to finally arrive at
Eq.(\ref{SM}).
\subsubsection{Expression for the Landauer conductance}
Now we substitute Eq.(\ref{SM}) in Eq.(\ref{curr}) with
$S_{\alpha\beta}(t,t')$ given by Eq.(\ref{SCM}) to get for the {\it linear
dc response} current $I_{0}=\overline{I_{\rm lin}(t)}$:
\begin{eqnarray}
\label{curr1}
I_{\rm lin}(t)&=& -v_{F}^{-2}\,U \int dt_{1}\int
dt_{2}\, [-i(t_{1}-t_{2})\,\hat{f}(t_{1}-t_{2})]\\ \nonumber
&\times& {\rm Tr}\left\{(1+\Lambda)
W{\bf G}^{R}(t,t_{1})W^{\dagger}\Lambda W{\bf G}^{A}(t_{2},t)W^{\dagger}
\right\}-
\\ \nonumber
&+& i(v_{F}^{-1}/2)\,U\int dt_{1}\,[-i(t_{1}-t)\,\hat{f}(t_{1}-t)]\\
\nonumber
&\times& {\rm
Tr}\left\{(1+\Lambda)W[{\bf G}^{R}(t,t_{1})-{\bf
G}^{A}(t_{1},t)]W^{\dagger}
\right\},
\end{eqnarray}
where we have introduced the diagonal matrix $\Lambda$: 
\begin{equation}
\label{Lambda}
\Lambda_{\alpha\alpha}=\left\{\matrix{+1, & \alpha\in {\rm lead\;\; 1}\cr
-1, & \alpha\in {\rm lead\;\; 2}\cr
0, & {\rm otherwise}}
\right.
\end{equation}
Equation Eq.(\ref{curr1}) can be simplified using the identity:
\begin{eqnarray}
\label{ident}
& &{\bf G}^{R}(t,t')-{\bf G}^{A}(t,t')= -{\bf G}^{A}\left\{[{\bf
G}^{-1}]^{R}-[{\bf
G}^{-1}]^{A}\right\}{\bf G}^{R}\nonumber \\
&=& -2iv_{F}^{-1}\,\int
dt_{1}\,{\bf
G}^{A}(t,t_{1})\,W^{\dagger}W\,
{\bf G}^{R}(t_{1},t') 
\end{eqnarray}
which is obtained with the help of Eq.(\ref{ImH}).

In order to obtain the dc $I_{0}$ one has to average Eq.(\ref{curr1})
over $t$ within the observation time ${\cal T}\rightarrow\infty$. This
means an additional integration over $t$. Then using
the fact that $t\hat{f}(t)$ is an even function one can replace
${\bf G}^{A}(t_{1},t)$ by ${\bf G}^{A}(t,t_{1})$ in the second
term of Eq.(\ref{curr1}). After that Eq.(\ref{ident}) can be applied to
yield:
\begin{eqnarray}
\label{LandF}
& & g_{\rm Land}= v_{F}^{-2}\,\int_{-{\cal T}/2}^{+{\cal
T}/2}\frac{dt}{{\cal T}}\int dt_{1}\int
d\eta\, [-i\eta\,\hat{f}(\eta)]
\nonumber\\   
&\times& 
{\rm Tr}\left\{-W^{\dagger}(1+\Lambda)\,
W{\bf G}^{R}(t,t_{1})W^{\dagger}\Lambda W{\bf
G}^{A}(t_{1}+\eta,t)
\right.
\nonumber \\
&+&\left.
W^{\dagger}W
{\bf G}^{R}(t,t_{1})W^{\dagger}(1+\Lambda)W{\bf
G}^{A}(t_{1}+\eta,t)
\right\}. 
\end{eqnarray}
Finally we assume  that the $2M\times N$ coupling matrix $W_{\alpha n}$ 
has only 2M nonzero matrix elements, those with
$\alpha = n$;
all these elements are  taken equal to W.
Then Eq.(\ref{LandF}) takes the form:
\widetext
\Lrule
\begin{equation}
\label{LF}
g_{\rm Land}=\frac{1}{2}\gamma^2 \int dt_{1}\int
d\eta\, [-i\eta\,\hat{f}(\eta)]\,M^{-2}\,{\rm Tr}\left\{\overline{{\bf
G}_{12}^{R}(t,t_{1})
{\bf G}_{21}^{A}(t_{1}+\eta,t)}+\overline{{\bf G}_{21}^{R}(t,t_{1})
{\bf G}_{12}^{A}(t_{1}+\eta,t)} \right\},
\end{equation}
\narrowtext
\noindent
where $\hat{f}(\eta)$ is the Fourier transform of $\tanh(E/2T)$; the
overline denotes the
average
over time $t$; $\gamma=2v_{F}^{-1}|W|^2 M$ is the electron escape rate,
$M$
is the number of channels in each lead and the subscripts 1 or 2
indicate that only sites connected to the first or the second lead
should be taken into account in the summation over matrix indices.

Using Eq.(\ref{ident}) one can \cite{VA} recast Eq.(\ref{LF}) as a sum of
two parts
$g_{\rm Land}= g_{1}+g_{2}$ where
\widetext
\Lrule
\begin{equation}
\label{g1}
g_{1}=\frac{i}{4}\gamma \int
d\eta\, [-i\eta\,\hat{f}(\eta)]\,M^{-1}\,{\rm tr}\left\{\overline{{\bf
G}^{R}(t+\eta,t)}-
\overline{{\bf G}^{A}(t+\eta,t)}\right\},
\end{equation}
and
\begin{equation}
\label{g2}
g_{2}=\frac{1}{4}\gamma^2 \int dt_{1}\int
d\eta\, [-i\eta\,\hat{f}(\eta)]\,M^{-2}\,{\rm
tr}\left\{\Lambda \overline{{\bf
G}^{R}(t,t_{1})\Lambda
{\bf G}^{A}(t_{1}+\eta,t)}\right\}.
\end{equation}
\Rrule
\narrowtext
\noindent
In Eqs.(\ref{g1}),(\ref{g2}) the symbol ${\rm tr}$ denotes the matrix
trace taken {\it only} over sites connected to leads and the matrix
$\Lambda$ in Eq.(\ref{g2}) is given by Eq.(\ref{Lambda}). 

The first contribution $g_{1}$ is proportional to the
local density of states in the regions of leads. It describes the
effect of electron escape in each lead {\it separately}. The second term
$g_{2}$
describes the mutual effect of {\it both} leads. It
is similar to the conventional term 
$\hat{\bf j}{\bf G}^{R}\hat{\bf j}{\bf G}^{A}$ in the Kubo conductance,
since ${\rm Tr}\hat{\bf j}={\rm tr}\Lambda=0$.

\subsection{Why there are no loose diffusons in the problem of Landauer
conductance under ac pumping}
We stress once again that the Landauer conductance is independent of the
Keldysh component of the electron Green's function in the dot.
This is already an indication that elecron {\it kinetics} inside the dot
under ac pumping is irrelevant for the Landauer conductance. In particular
this means that the electron energy distribution function   
$\hat{f}(\eta)$ in Eq.(\ref{LF}) {\it is not renormalized} in the way   
similar to Eq.(\ref{Fren}).

Formally this follows from the difference in the structure of the {\it
retarded-advanced junctions}. Let us compare the retarded-advanced
junction Eq.(\ref{RAj}) and that in Eq.(\ref{g2}). The difference is that
Eq.(\ref{RAj}) contains the electron interaction with the
{\it ac pumping
field} $\hat{{\bf r}}\partial_{t_{1}}{\bf A}
(t_{1})-\hat{{\bf
r}}\partial_{t_{1}}{\bf A}(t_{1}+\eta)$
or 
$\hat{{\bf j}}{\bf A}
(t_{1})-\hat{{\bf
j}}{\bf A}(t_{1}+\eta)$.
At the same time
Eq.(\ref{g2}) contains $-i\eta\Lambda\propto \Lambda\,[{\bf A}_{0}(t_{1})-
{\bf A}_{0}(t_{1}+\eta)]$
which
stems from the interaction with the
{\it dc field} ${\bf
A}_{0}(t)=-{\bf E}_{0}t$ that causes the chemical potential difference
$U$. Correspondingly, the expression for the loose diffuson 
analogous to Eq.(\ref{ldif}) is proportional \cite{symm} to 
$\Lambda_{\mu 0}=\sum_{m}\Phi_{\mu}(m)\Lambda_{m m}\Phi_{0}(m)
\propto {\rm tr}\Lambda=0$ (instead of ${\bf r}_{\mu
0}$). An expansion of ${\bf G}^{R,A}$ in powers
of the ac
pumping field 
analogous to that leading to the triangle in Fig.4 does not
help. Because the retarded-advanced junction in Eq.(\ref{g2}) does not
contain the ac field the
corresponding triangle is {\it linear}
in
the ac field   
and vanishes
after averaging over time $t$ [cf. Eq.(\ref{Gamma})].
Thus the loose diffuson corresponding to
Eq.(\ref{g2}) vanishes in the same way as the one that corresponds to the
conventional term $\hat{{\bf j}} {\bf G}^{R}\hat{{\bf j}}{\bf G}^{A}$ in
the Kubo
conductance. The structure of Eq.(\ref{g1}) does not allow to build the
loose diffuson because it does not contain the retarded-advanced junction
al all.

Since the heating effect is associated with the loose diffusons the above
arguments allow one to 
conclude \cite{AleinPr} that the  {\it Landauer conductance}
of
non-interacting electrons 
is {\it not} sensitive
to heating.
\subsection{Sensitivity of the Kubo conductance to heating}
However, this is not true for the Kubo conductance which {\it is
sensitive} to heating. 

Consider an experimental situation  shown in Fig.10b. Here the
electric field $E_{0}$ is
produced by the linear in time component of the magnetic flux piercing the
mesoscopic ring $\phi_{0}= L A_{0}= -L E_{0}t$. 
The ac pumping is produced by the oscillating part of the magnetic flux 
$\phi_{\rm ac}(t)=A(t) L$.  
In the framework of the dynamical approach the steady-state regime is only
reachable if the mesoscopic system is connected to the reservoir by a 
{\it passive lead}. It supports no net current but results
in
the
particle exchange between the mesoscopic system and the reservoir
necessary to remove heat produced by the ac pumping. Phenomenologically,
the effect of the passive lead is described by the escape rate $\gamma$.
This geometry corresponds exactly to the formalism developed in Sec.II and
Sec.III, where
\begin{equation}
\label{intK}
{\cal H}_{\rm e-f}(t)={\cal H}_{0}+{\cal H}_{1}=\hat{{\bf j}}{\bf E}_{0}t
- \hat{{\bf j}}{\bf A}(t).
\end{equation}
Now each ray in Fig.2 represents a sum ${\cal H}_{0}+{\cal H}_{1}$.  
Since we are interested in the linear response to ${\cal H}_{0}$, one ray
is special: it corresponds to ${\cal H}_{0}$ while all other rays
correspond to ${\cal H}_{1}$. Then using
Eqs.(\ref{I-expansion})-(\ref{ra}) one obtains $I_{0}=\overline{I_{\rm
lin}(t)}=[\sigma_{1}+\sigma_{2}+\sigma_{3}]\,E_{0}$, where:
\widetext
\Lrule
\begin{eqnarray}
\label{sigma1}
\sigma_{1}= \int dt_{1}\int d\eta\,\hat{f}(\eta)\,{\rm
Tr}\left\{
\hat{{\bf j}} \overline{{\bf G}^{R}(t,t_{1})\hat{{\bf j}}\,[-it_{1}] {\bf
G}^{R}(t_{1},t-\eta)}-\hat{{\bf j}} \overline{{\bf
G}^{A}(t+\eta,t_{1})\hat{{\bf
j}}\,[-it_{1}] {\bf
G}^{A}(t_{1},t)}\right\},
\end{eqnarray}
\begin{eqnarray}
\label{sigma2}
\sigma_{2}= \int dt_{1}\int d\eta\,[-i\eta\hat{f}(\eta)]\,{\rm 
Tr}\left\{
\hat{{\bf j}} \overline{{\bf G}^{R}(t,t_{1})\hat{{\bf j}} {\bf
G}^{A}(t_{1}+\eta,t)}\right\}, 
\end{eqnarray} 
\begin{eqnarray}
\label{sigma3}
\sigma_{3}=\int dt_{1}\,{\rm Tr} \left\{
\hat{{\bf j}} \overline{{\bf G}^{(a)}(t,t_{1})\hat{{\bf j}}\,[-it_{1}]{\bf
G}^{A}(t_{1},t)}+\hat{{\bf j}}\overline{{\bf
G}^{R}(t,t_{1})\hat{{\bf j}}\,[-it_{1}]{\bf G}^{(a)}(t_{1},t)}\right\}.
\end{eqnarray}
Here ${\bf G}^{R,A}(t,t')$ are Green's functions in the mesoscopic system
subject to ac pumping and
\begin{equation}
\label{Ganom}
{\bf G}^{(a)}(t,t')=\int dt''\int d\eta\,\hat{f}(\eta)
\,{\bf
G}^{R}(t,t''+\eta)[{\cal H}_{1}(t''+\eta)-
{\cal H}_{1}(t'')]{\bf
G}^{A}(t'',t').
\end{equation}
\Rrule
\narrowtext
\noindent
The first two parts Eqs.(\ref{sigma1}),(\ref{sigma2}) do not allow to
build the loose diffuson for exactly the same reasons as
Eqs.(\ref{g1}),(\ref{g2}) for the Landauer conductance. However, the
`anomalous' Green's function
$G^{(a)}$ in Eq.(\ref{sigma3}) contains the retarded-advanced junction of
the same structure as we discussed in Sec.IV which does allow to
build the loose diffuson. This is the part where the heating effects
oridinate from.

One can see the direct analogy with the situation discussed in Sec.Vb.
The role played by the current $I^{(2)}$, Eq.(\ref{I2}), is now played by
the part of conductance $\sigma_{3}$. The parent diagrams for the problem
of conductance fluctuations are the usual two-diffuson and two-cooperon
diagrams considered in Refs.\cite{ALS}. All the daughter-diagrams 
with one or two additional loose diffusons can be obtained from the
disorder averages $\langle\sigma_{1,2}\sigma_{3}\rangle$ and
$\langle \sigma_{3}^2 \rangle$, respectively. It is not difficult to show
that all of
them again lead to the renormalization of the energy distribution function
given by Eq.(\ref{Fren}).

\subsection{The difference between the Landauer and the Kubo conductance}
Let us discuss the difference between the experimental situations
described by the
Landauer and the Kubo conductances. The
principal difference between them is that the quantity we call
the Landauer conductance is the linear current response to the
variations
of the {\it parameters of the electron reservoirs} (the chemical potential
difference) while the Kubo conductance is the linear current response
to the variation of the {\it system's Hamiltonian} (the term ${\cal
H}_{0}$ in Eq.(\ref{intK})). We stress that the
difference lies in the {\it physical situations} and not in the
{\it methods of
description}. For instance, the {\it photovoltaic effect} 
can be considered both in the framework of the scattering matrix approach
\cite{VAA} and in the nonlinear response approach similar to the one we
used in Sec.V for the
persistent current fluctuations. In both cases the result is expressed in
terms of the {\it renormalized} electron energy distribution
Eq.(\ref{Fren}). The reason is that the photovoltaic current is a
non-linear response to the variation of the system's Hamitlonian
(${\cal H}_{1}$ in Eq.(\ref{intK})) caused by
the ac pumping field which should be treated in the same way as the
variation ${\cal H}_{0}$ in Eq.(\ref{intK}) that causes the Kubo
conductance. 
That is why in terms of the sensitivity to heating the photovoltaic effect
is similar to the Kubo conductance
and drastically different from the Landauer conductance. 

We note that the key difference between these two situations examplified
by Fig.10a and Fig.10b is that the initial
density matrix at $t=-\infty$ {\it prior to switching on the interaction
with external ac field} (which is a starting point in the Keldysh
formalism
\cite{keld}), is already {\it non-equilibrium} in the problem of Landauer
conductance [Fig.10a], since it involves {\it two} Fermi-distributions
with
different chemical potentials. At the same time in the situation
represented by Fig.10b the initial density matrix is equilibrium.

\section{Conclusion}
In the present paper we have considered a general formalism of the
dynamical approach to nonlinear response of mesoscopic systems. For
completeness of presentation and for tutorial purposes we have shown how
causality principle leads to the 
powerful machinery of {\it analytical continuation} \cite{elias} which
is most conveniently realized in the matrix algebra of the {\it Keldysh
technique} \cite{keld,LO-1986,RS-1986}. The principal goal of the paper 
was to demonstrate the capability of the dynamical approach and its
apparent limitations. We have shown that the diagrammatic technique
developed in Ref.\cite{GLK-1979,H-1981} automatically describes the
electron diffusion in the energy space under the action of ac pumping. 
We considered few simple examples of how the renormalization of the
electron
energy distribution occurs  due to the {\it
loose diffusons} and have
demonstrated its equivalence to the solution to the kinetic equation.

We have shown that the quadratic response in the {\it closed} mesoscopic
systems is {\it singular} in the dynamical approach because of the
singularity of the {\it loose diffusons} in the quadratic in
the pump field
approximation. Yet in contrast to 
Ref.\cite{Kop} this does not lead to an {\it infinite} dc current arising
under ac pumping, as the quadratic in ${\bf A}$ approximation breaks down
well before the singularity develops itself. The situation is somewhat
opposite to that described in Ref.\cite{Kop}. The ensemble-averaged dc
current does not depend on the singular loose diffusons whatsoever
\cite{average} and thus is finite anyway. 
The mesoscopic fluctuations of the dc
current in an isolated ring are {\it zero} because the corresponding
effective temperature $T_{*}\rightarrow\infty$ in the absence of
dissipation or an electron escape. This clearly shows that the dynamical
approach based on the `minimal model' of non-interacting electrons in an
impurity potential intercating with the external classical field is {\it
insufficient} for describing the closed systems.  

However, this model is reasonable for  {\it open} mesoscopic systems
connected with the electron reservoir by massive leads.
We have carefully studied the case of the Landauer conductance in a
quantum dot under ac pumping, since here one can see the danger of
{\it ad hoc} replacement of the bare electron energy distribution by the
renormalized one with the simultaneous deletion of all the diagrams with
the loose diffusons. The correct diagrammatic analysis shows that the
diagrams with the loose diffusons do not arise in this problem, and the
electron energy distribution that enter e.g. the variance of conductance
fluctuations stays unrenormalized by heating. At the same time, the same type of
diagrammatic analysis shows that the  variance of the Kubo conductance
is sensitive to the renormalization of electron energy distribution.
This sets yet another borderline between these two formulations of the
problem
of conductance and shows their deep physical difference and significance
of the experimental geometry for conductance measurements.

Though the results presented in the paper are purely perturbative, there
is a bridge to non-perturbative schemes. The obvious extension of the
present theory is the {time-dependent random matrix theory} (TRMT) which
applies
to the problem of quantum dot under ac pumping. We found that in contrast
to the equilibrium or the linear response theories, there are {\it two}
different TRMT formulations Eq.(\ref{RMT1}) and Eq.(\ref{RMT2}), one for
the
adiabatic pumping with the typical frequency $\omega$ smaller than the
Thouless
energy $E_{c}$ and another one for the case of high frequencies
$\omega\gg E_{c}$. The corresponding matrix Hamiltonians have different
symmetry: they are real symmetric for the adiabatic case and contain an
imaginary anti-symmetric time-dependent part (with zero time-average) in
the case of high frequencies.

Another obvious extension is the functional formulation in terms of the
{\it non-linear sigma-model} \cite{Kam}. In this connection we note that
in
all field theories with the
correct vacuum, the loose propagators may not arise. In the particular
formulation of the non-linear sigma-model based on the Keldysh formalism
the
electron energy
distribution is a part of the vacuum solution. However, the example of
the Landauer conductance (that depends on the bare energy distribution
function ) and the Kubo conductance or
the photovoltaic effect \cite{VAA} (that depend on the renormalized
energy distribution function) shows that the vacuum could be
non-unique
or containing different sectors which enter in a different way in one or
another observable. 


Finally we mention the  effect of electron-electron interaction.
It seems plausible that in the presence of an electron-electron
interaction the Landauer conductance will be sensitive to heating.
Another effect of electron interaction  
is  the {\it inelastic electron scattering}. In our dynamical
approach we have neglected this effect at all. In open mesoscopic
systems this has resulted in the
electron energy distribution Eq.(\ref{Fren}) of a very peculiar form.
The characteristic feature of this distribution is that it depends not
only on the {\it effective temperature} $T_{*}$ but also on the bath
temperature $T\ll T_{*}$. The consequence of this two-parameter dependence
takes its extreme form in the temperature dependence of the persistent
current or photovoltaic current fluctuations. In particular,
Eq.(\ref{EXAMPLE}) shows how the {\it separation of parameters} occurs in
the variance of the persistent current fluctuations under ac pumping. The
parameter $T_{*}$ turns out to determine only the {\it pre-factor} in
front of the
function
that depends on the bath temperature $T$. Such a behavior is only possible
if electron-electron interaction is neglected. Inelastic processes
due to electron-electron interaction are always favorable to the
Fermi-distribution function with some effective temperature $T_{el}$.
They
work to
diminish all the deviations from the Fermi-distribution and should
certainly reduce
the dependence of the persistent current
fluctuations on the bath temperature. This (or similar
temperature dependences for the
photovoltaic effect) can be a useful tool for the experimental
investigation
of electron-electron interaction in an open quantum dot.

We are grateful to I.L.Aleiner, 
A. Kamenev and I.V.Lerner for illuminating discussions. V.I.Yudson
appreciates the support from the
Ministry of  Education of Japan
(Mombusho grant 12640338), the Russian Ministry of Science
("Nanostructures"
program),
and RFBR (grant No.98-02-16062), and  acknowledges the hospitality of  the
ICTP in Trieste, where a part of  the work was done.

\widetext

\begin{references}
\bibitem{average} The 
cancellation does not hold for the ensemble-average value which
for non-interacting electrons is  expressed in terms of
$\int_{-\infty}^{+\infty}dE \,\partial
F(E)/\partial E=i\eta \hat{F}(\eta)|_{\eta=0}$. This quantity at any
finite
$\gamma$ is equal to its equilibrium value of 2, since all 
loose diffusons vanish at $\eta=0$.
\bibitem{outgoing} The matrix Green's function
$\underline{G}_{--,\alpha\beta}(t,t';x-x')$ for the outgoing electrons
still depends only on the difference of coordinates but in the presence
of ac pumping inside the dot is no longer a function of the difference of
time.
\end{references}
\end{document}